\RequirePackage{fix-cm}
\documentclass[twocolumn]{svjour3}          
\smartqed  
\usepackage{graphicx}
\usepackage{cite}
\usepackage{amsmath,amssymb,amsfonts}
\usepackage{algorithmic}
\usepackage{graphicx}
\usepackage{subcaption}
\usepackage{array}
\usepackage{float}
\usepackage{multirow}
\usepackage{textcomp}
\usepackage{xcolor}
\usepackage[utf8]{inputenc}
\usepackage{csquotes}
\usepackage{comment}
\usepackage[acronym]{glossaries}
\usepackage{svg}
\usepackage{booktabs}
\usepackage{tabularx}
\usepackage{color}
\usepackage[figure,table]{totalcount}
\usepackage[symbol]{footmisc}
\usepackage{tablefootnote}

\newacronym{ROI}{ROI}{Region of Interest}
\newacronym{IARC}{IARC}{International Agency for Research on Cancer}
\newacronym{SSDL}{SSDL}{Semi-supervised Deep Learning}
\newacronym{CAD}{CAD}{Computer Aided Diagnosis}
\newacronym{PBC}{PBC}{Pseudo-label based Balance Correction}
\newacronym{CBIS-DDSM}{CBIS-DDSM}{Curated Breast Imaging Subset of Digital Database for Screening Mammography}
\newacronym{TC}{TC}{true certainty}
\newacronym{FC}{FC}{false certainty}
\newacronym{TU}{TU}{true uncertainty}
\newacronym{FU}{FU}{false uncertainty}
\newacronym{TCIA}{TCIA}{The Cancer Imaging Archive}
\newacronym{TP}{TP}{true positives}
\newacronym{TN}{TN}{true negatives}
\newacronym{FP}{FP}{false positives}
\newacronym{FN}{FN}{false negatives}
\newacronym{AUROC}{AUROC}{Area Under the Receiver Operating Characteristic Curve}
\newacronym{DeDiMs}{DeDiMs}{Deep data set Dissimilarity Measures}
\newacronym{CC}{CC}{Craniocaudal}
\newacronym{MLO}{MLO}{Mediolateral Oblique}
\newacronym{ACR}{ACR}{American College of Radiology}

\renewcommand{\thefootnote}{\arabic{footnote}}

\journalname{Medical \& Biological Engineering \& Computing}

\begin{document}
\sloppy
\title{A Real Use Case of Semi-Supervised Learning for Mammogram Classification in a Local Clinic of Costa Rica
}


\author{ Sa\'ul Calder\'on-Ram\'irez         \and Diego Murillo-Hern\'andez \and  Kevin Rojas-Salazar \and David Elizondo \and Shengxiang Yang \and Armaghan Moemeni \and Miguel Molina-Cabello
}


\institute{Sa\'ul Calder\'on-Ram\'irez \and David Elizondo \and Shengxiang Yang \at
              Centre for Computational Intelligence (CCI), De Montfort University, United Kingdom
           \and
           Sa\'ul Calder\'on-Ram\'irez \and Diego Murillo-Hern\'andez  \and Kevin Rojas-Salazar \at PARMA Research Group, Instituto Tecnol\'ogico de Costa Rica, Costa Rica
            \and
           Diego Murillo-Hern\'andez  \and Kevin Rojas-Salazar \at
              School of Computing Engineering, Instituto Tecnol\'ogico de Costa Rica, Costa Rica
              \and
            Armaghan Moemeni \at School of Computer Science, University of Nottingham, United Kingdom
            \and
            Miguel Molina-Cabello \at Department of Computer Languages and Computer Science. University of M\'alaga, Spain
           \\
              \\
              \email{sacalderon@itcr.ac.cr, diemurillo@ic-itcr.ac.cr, kevin.rojas7@estudiantec.cr, elizondo@dmu.ac.uk, syang@dmu.ac.uk, armaghan.moemeni@nottingham.ac.uk, miguelangel@lcc.uma.es} 
}

\date{Received: date / Accepted: date}

\maketitle
{\raggedright{\textbf{Total Tables:} \totaltables \\ \textbf{Total Figures:} \totalfigures} \\[1pc]} 

\begin{abstract}
The implementation of deep learning based computer aided diagnosis systems for the classification of mammogram images can help in improving the accuracy, reliability, and cost of diagnosing patients. However, training a deep learning model requires a considerable amount of labeled images, which can be expensive to obtain as time and effort from clinical practitioners is required. A number of publicly available datasets have been built with data from different hospitals and clinics. However, using models trained on these datasets for later work on images sampled from a different hospital or clinic might result in lower performance. This is due to the distribution mismatch of the datasets, which include different patient populations and image acquisition protocols. The scarcity of labeled data can also bring a challenge towards the application of transfer learning with models trained using these source datasets. In this work, a real world scenario is evaluated where a novel target dataset sampled from a private Costa Rican clinic is used, with few labels and heavily imbalanced data. The use of two popular and publicly available datasets (INbreast and CBIS-DDSM) as source data, to train and test the models on the novel target dataset, is evaluated. The use of the semi-supervised deep learning approach known as MixMatch, to leverage the usage of unlabeled data from the target dataset, is proposed and evaluated. In the tests, the performance of models is extensively measured, using different metrics to assess the performance of a classifier under heavy data imbalance conditions. It is shown that the use of semi-supervised deep learning combined with fine-tuning can provide a meaningful advantage when using scarce labeled observations. We make available the novel dataset for the benefit of the community. 

\end{abstract}

\section{Introduction}
    
    Breast cancer is one of the leading causes of death in women around the world \cite{wild2020world}. Nonetheless, it is widely known that diagnosing a malign breast tumor in its early stages can increase treatment effectiveness \cite{american2019breast}. In many situations, an early diagnostic can increase survival probability significantly. 
    
    Deep learning has extensively been explored and implemented as an approach to develop \gls{CAD} systems using medical imaging \cite{calderon2018assessing,bermudez2020first,calvo2019assessing,alfaro2019brief,calderon2020correcting}. Such systems have the potential of highly improving the diagnosis and further treatment of patients. For mammogram analysis, different deep learning architectures have been proposed, for either binary classification, BI-RADS based multi-class classification, or segmentation of regions of interest \cite{hamidinekoo2018deep,abdelhafiz2019deep}.   However, the usage of deep learning in the medical domain faces the challenge of scarce labeled data.
    
    In this work, we explore the following setting: take a specific clinic or hospital  to deploy a deep learning model (target hospital/clinic). Therefore, data sampled from such hospital must be used for evaluation purposes. Additionally, a small number of labeled observations  sampled from such target hospital/clinic might be available. Furthermore, different datasets sampled from other hospitals or clinics might also be available.  The notation of such experimental settings can be formalised as follows:

\begin{itemize}
    \item Target labeled dataset $D^l_t$: A small number of labeled observations $n^l_t$ might be available which can be used for training/fine-tuning the model. 
    \item Source labeled dataset $D^l_s$: Different data sources of data sampled in different hospitals/clinics might be used. Usually these datasets have a large number of labeled observations, thus $n^l_t < n^l_s$.
    \item Target unlabeled dataset $D^u_t$:  A larger number of unlabeled observations $n^u_t$ might be available and can also be used for training/fine-tuning the model. As unlabeled data is cheaper to obtain, it can often be found that $n^l_t < n^u_t$.
    \item Source unlabeled dataset $D^u_s$: Similarly to the aforementioned case, more source unlabeled observations might be available when compared to the number of source labeled observations, thus $n^l_s < n^l_s$. 
\end{itemize}

In this work, the usage of both transfer and semi-supervised learning using two different source datasets is explored: INbreast ($D^l_{s,\textrm{IN}}$) \cite{inbreast} and CBIS-DDSM ($D^l_{s,\textrm{DDSM}}$) \cite{CBIS-DDSM-citation}. The target dataset was obtained from the Costa Rican medical private clinic Im\'agenes M\'edicas Dr. Chavarr\'ia Estrada (hereafter referred as  $D^{l}_{t,\textrm{CR}}$). The aim of this research is to experiment the effectiveness of fine-tuning deep learning models in a semi-supervised fashion (using both $D^u_t$ and $D^l_t$), performing transfer learning from models trained with the source datasets $D^l_{s,\textrm{DDSM}}$ and $D^l_{s,\textrm{IN}}$. For this study, the usage of unlabeled data from other source datasets was avoided, as it has been reported that it might decrease the performance of a \gls{SSDL} model \cite{calderon2020mixmood,calderon2021more}.   In this work, we use MixMatch as a semi-supervised learning approach \cite{berthelot2019mixmatch}, given previously positive results reported for this approach in medical imaging \cite{calderon2020dealing,calderon2021improving}.
    
\section{State of the Art}

\subsection{Transfer learning and data augmentation for mammogram classification}

\gls{CAD} of breast cancer via mammogram image classification  has been widely studied in the literature. Authors in \cite{abdelhafiz2019deep} present a survey of the state of the art in the application of deep learning in the analysis of mammography images for the early detection of breast cancer. The authors summarize open challenges and best practices to follow when dealing with mammogram analysis using deep learning. One of the most frequent short-comings of implementing deep learning for mammogram analysis in a target clinic/hospital is the lack of labeled training data \cite{abdelhafiz2019deep}. Labeling medical images can be particularly expensive, as trained professionals are needed to carry out such specialized tasks \cite{SUN20174}. To overcome this challenge,  a number  mammogram datasets are publicly available. However, different patient populations and image acquisition protocols can limit and hinder the performance of the final model using the target data \cite{CBIS-DDSM-publication}. 

Two of the most common approaches to tackle the problem of labeled data scarcity are transfer learning and data augmentation \cite{hamidinekoo2018deep,abdelhafiz2019deep}. Using pre-trained model parameters from more general tasks often improve the model's performance. Authors in \cite{falconi2020transfer} experimented with the multi-class classification of mammograms using transfer learning from ImageNet. Similarly, authors in \cite{PARDAMEAN2018400} observed encouraging results in the classification of mammograms when using transfer learning from a chest X-ray dataset of patients with pneumonia.

Applying transfer learning with models trained with observations from the same domain is intuitively an interesting approach. Authors in \cite{alkhaleefah2020double} carried out an exhaustive research for improving the performance of deep learning models in the binary classification of mammogram anomalies by using features previously learned from different mammogram datasets. Authors in \cite{shen2019deep} also experimented with transfer learning from mammogram datasets for the detection and classification of anomalies in mammogram images. For these cases the more specific term \enquote{domain adaptation} can be used, as although images from different datasets can be visually and semantically similar, their distributions might be significantly different, as explained in \cite{CHEPLYGINA2019280,Tajbakhsh2017}.

As previously mentioned, data augmentation is also an effective approach to tackle data scarcity \cite{abdelhafiz2019deep}. Simple  augmentations by applying common image transformations like image rotations and flips can improve results \cite{levy2016breast}. In previous works, more sophisticated and domain-specific data augmentation techniques have been developed \cite{domingues2018bi}. Authors in  \cite{castro2018elastic} obtained positive results  by implementing elastic deformations for mammogram images, simulating possible different views of the same breast. The augmentation of training data has also been recently achieved by creating artificial observations with generative deep learning models \cite{wu2020synthesizing, korkinof2019highresolution}.

\subsection{Semi-supervised learning for medical imaging and mammogram analysis}

Another approach to deal with small labeled datasets is the usage of \gls{SSDL}, which leverages unlabeled data to improve the model's performance  \cite{CHEPLYGINA2019280}. In recent years, the usage of the cheaper and larger unlabeled datasets for training deep learning models has proven to be a viable option for handling the lack of labeled data, as well as improving the performance of models \cite{calderon2020dealing,calderon2020correcting}. Authors in \cite{CHEPLYGINA2019280} present a survey of recent literature of semi-supervised learning approaches for medical imaging. The survey shows how  unlabeled datasets  have been used for improving model training in brain tumor segmentation, detection of vascular lesions, and prostate cancer detection. More recently, the usage of unlabeled data with semi-supervised deep learning has proven to give positive results in the detection of COVID-19 in chest x-ray images \cite{calderon2020dealing,calderon2020correcting}.

However, research on \gls{SSDL} approaches for mammogram analysis is still limited. In \cite{9350835} the authors propose a new semi-supervised architecture for convolutional neural networks, designed to extract information from multiple views of masses from mammogram images for their binary classification. In \cite{BAKALO202115} a semi-supervised setup is proposed for the joint use of weakly labeled data with fully labeled data of mammogram regions in the detection and classification of anomalies. Authors in \cite{SUN20174} also proposed a semi-supervised approach based on graphs and convolutional neural networks for the classification of anomalies in mammograms. However, from our knowledge few authors in the literature deal with the classification of mammograms using less expensive whole-image labels only. In \cite{calderon2021improving} the MixMatch approach was tested to improve the accuracy and predictive uncertainty of models applied to the binary classification of whole mammogram images. A target hospital or clinic might not have lower level labels available, to fine-tune and test a deep learning model.  

As previously mentioned, analysis of mammograms includes lower level tasks such as: segmentation and detection of anomalies, the higher abstraction of level tasks, the binary classification of images (malign findings with no/benign findings) \cite{hamidinekoo2018deep, abdelhafiz2019deep}. It may also include  multi-class classification, for instance using the BI-RADS standard \cite{domingues2018bi}. As such, different levels of annotations in the data might be needed for lower level tasks, like pixel-level annotations of the \gls{ROI}. When using transfer learning to leverage information from thoroughly annotated source datasets for lower level tasks, fine-tuning on the target data might still be needed  \cite{CHEPLYGINA2019280}. This, similar degrees of annotations would be preferable in the target dataset as well. Therefore, the need to use target data to train or fine-tune a model  makes the use of unlabeled data an interesting alternative. Different image acquisition protocols and patient distribution sampled in a  dataset source is a frequent real-life scenario that increases the need of model fine-tuning.

\subsection{\gls{SSDL} with MixMatch}
In this work, the MixMatch method is used as the semi-supervised learning approach for training models with unlabeled data. This is novel \gls{SSDL} method, presented by the authors in \cite{berthelot2019mixmatch} has shown important accuracy gains against previous \gls{SSDL} frameworks. It is mainly based on the use of pseudo-labels, unsupervised regularization and data augmentation. The following corresponds to a brief description of the method.

\gls{SSDL} makes use  of labeled and unlabeled observations $X_l$, $X_u$ respectively. MixMatch implements data augmentation with affine transformations on both datasets. Pseudo-labels are then generated for each unlabeled observation, sharpening the average of the predictions of a model on each of its augmented \enquote{versions}. This results in the set $\tilde{Y}$ of pseudo-labels for observations of $X_u$. Similarly, the set $Y_l$ can be used to represent the labels of observations in $X_l$.

Further data augmentation is applied to the datasets $S_l$ and $\tilde{S}_u$, with $S_l = (X_l, Y_l)$ and $\tilde{S}_u = (X_u, \tilde{Y})$, by using linear interpolation of the data with the MixUp algorithm, as mentioned in \cite{berthelot2019mixmatch}. This way, the sets of augmented data $\tilde{S}_{u}'$ and $S_{l}'$ are obtained and finally used to train a model by minimizing the compound loss function shown in equation \ref{eq_mixmatch}.

\begin{equation}
    \label{eq_mixmatch}
    \begin{split}
        \mathcal{L}(S,\theta) = \sum_{(x_i,y_i)\in S_{l}'} \mathcal{L}_l (\theta,x_i,y_i) + \\
        \gamma r(\tau) \sum_{(x_j,\tilde{y}_j)\in \tilde{S}_{u}'} \mathcal{L}_u (\theta,x_j,\tilde{y}_j)
    \end{split}
    \end{equation}

This loss function is formed by the respective supervised and unsupervised loss terms $\mathcal{L}_l$ and $\mathcal{L}_u$. In this work, the former is implemented as a cross-entropy loss, while the latter is implemented as an Euclidean distance, with the regularization coefficient $\gamma$ and the rampup function $r(\tau) = \tau / 3000$, as recommended in \cite{calderon2020dealing}.

\subsection{Class imbalance correction}
A major factor that must be taken into account in the process of implementing a model for classification tasks, specially in the medical domain, is the distribution of classes in a dataset \cite{abdelhafiz2019deep}. For medical conditions, it is common for observations depicting a disease or a \enquote{positive} case, to be fairly less frequent in comparison to normal or healthy observations \cite{calderon2020correcting}. Training a model with imbalanced data can lead to the final model being biased towards the majority classes, while ignoring the minorities.

Multiple approaches to tackle the problem of imbalanced class distributions in datasets can be found in the literature \cite{calderon2020correcting}. Two of the most straightforward techniques used include under-sampling and over-sampling \cite{7727770}. These techniques, although fairly simple and intuitive, might not prove to be the best choice, as they can lead respectively to information loss and over fitting \cite{7727770}. Other common approaches used towards imbalanced class distributions in datasets involve the so called \enquote{cost sensitive learning} \cite{7727770}. One implementation of this approach is to give weights to each class inside the cross-entropy loss function to correct for class imbalance. In the case of semi-supervised learning, authors in \cite{calderon2020correcting} proposed a similar technique called \gls{PBC}. This technique applies class-balance correction both to the labeled and unlabeled data in the MixMatch \gls{SSDL} approach. Given its reported positive results, we implement the class imabalance correction approach tested in \cite{calderon2020correcting} in our work.

\subsection{Classification Metrics for Imbalanced Data}
\label{Metrics}
Class-imbalanced  datasets and its impact on the implementation of classification models has long been a subject of study in the literature \cite{kubat1997addressing}. Using metrics that account for class imbalance is an important aspect, specially for \gls{CAD} systems used under real-life conditions.  The most frequent and  almost customary method for evaluating the classification performance of models consists in the traditional classification accuracy \cite{sokolova2006beyond}. It  is measured as follows for the case of binary classification, which is the focus of this work:

\begin{equation}
   \text{Accuracy} = \frac{\text{TP} + \text{TN}}{\text{TP} + \text{TN} + \text{FP} + \text{FN}}
    \label{eq:1}
\end{equation}

With \gls{TP}, \gls{TN}, \gls{FP} and \gls{FN} representing all elements in the traditional confusion matrix scheme for binary classification. Despite its wide usage, traditional accuracy is not an adequate metric for imbalanced test data settings \cite{akosa2017predictive}. As seen in Equation \ref{eq:1}, this metric does not take into account the possible differences between the distribution of both classes, and thus can mislead to optimistic results, as illustrated by authors in \cite{chicco2020advantages}. A simple example of this situation consists of a scenario where $90\%$ of observations from a dataset belong to the negative class and the remaining $10\%$ belong to the positive class. If a model predicts all observations as negative, it would have an overly optimistic accuracy of $90\%$. The author in \cite{akosa2017predictive} explains that in such cases, traditional accuracy might reveal more information about label distribution than the model's performance.
    
Basic and widely known classification metrics that also derive from the confusion matrix scheme are the recall, specificity, and precision \cite{akosa2017predictive}. These metrics offer more  information about the model's classification performance and have been used in the literature to provide more complete analysis in cases with imbalanced data settings \cite{akosa2017predictive, johnson2019deep}. These are defined as:

\begin{equation}
   \text{Recall} = \frac{\text{TP}}{\text{TP} + \text{FN}}
    \label{eq:2}
\end{equation}

\begin{equation}
   \text{Specificity} = \frac{\text{TN}}{\text{TN} + \text{FP}}
    \label{eq:3}
\end{equation}

\begin{equation}
   \text{Precision} = \frac{\text{TP}}{\text{TP} + \text{FP}}
    \label{eq:4}
\end{equation}

Recall can be interpreted as the true positive rate, or accuracy for the positive class. It is the complement of the false negative rate. In a similar fashion, specificity can be interpreted as the true negative rate or accuracy for the negative class. It is the complement of the false positive rate. As for the precision, it indicates the probability of a model's positive predictions being correct. Precision, sensitivity and specificity measures provide values in the interval $[0,1]$, where higher is better. While these metrics can be studied individually to analyse different dimensions of the performance of a model, other metrics can be used to summarize them into a single score or value.  As discussed by the authors in \cite{chicco2020advantages},  currently there is no  consensus in the  machine learning community on the ideal classification metric to use, specially in cases with imbalanced data.

Two of the most widely used classification metrics, besides traditional accuracy, are the F-1 Score and \gls{AUROC}. These metrics are commonly used in contexts prone to data imbalance, such as information retrieval \cite{powers2015f} and the medical domain \cite{sokolova2006beyond}, although they are not always adequate for such cases \cite{maratea2014adjusted}. The F-1 score corresponds to the  harmonic mean between recall and precision. As such, it focuses on the positive class and does not take into account the \gls{TN} predictions of a model, as seen on Equation \ref{eq:5}. This metric is most useful in contexts where the main focus of a problem is the positive class, and the detection of the negative class is less relevant \cite{sokolova2006beyond}. It offers a balanced score of the rate of true positives (recall) and the rate of correctly predicted positives (precision).

Nevertheless, multiple works and studies point out the deficiencies of this metric and discourage its use as a standalone measure for the classification performance of a model \cite{powers2015f, chicco2020advantages, maratea2014adjusted, forman2010apples}, specially in cases of high class imbalance. Namely, one of the problems commonly pointed out is the fact that the F-1 Score weights the \gls{FP} the same as the \gls{FN}, as it can be seen in Equation \ref{eq:5}. To address this short-coming in imbalanced data scenarios is the F-2 score \cite{devarriya2020unbalanced},  measured as follows:

\begin{equation}
    \label{eq:5}
    \begin{split}
        \text{F$\beta$ Score} = (1 + \beta^{2})\cdot \frac{\text{Precision}\cdot\text{Recall}}{(\beta^{2}\cdot\text{ Precision}) + \text{Recall}} \\
        = \frac{(1 + \beta^{2})\cdot\text{TN}}{(1 + \beta^{2})\cdot\text{TP} + \beta^{2}\cdot\text{FN} + \text{FP}}
    \end{split}
\end{equation}

The F-1 Score and the F-2 Score can be generalised as the F$\beta$ Score, with $\beta = 1$ and $\beta = 2$ respectively.

The \gls{AUROC} is another single score metric that summarises the trade-off between the rate of true positives and the rate of false positives given multiple decision thresholds for the classification performance of a model. It provides a deeper insight of the model's behavior, when compared to the accuracy. However, it still faces many problems that are pointed out by a  number of authors in the literature \cite{chicco2020advantages, berrar2012caveats, hand2009measuring}, some related to the impact of highly imbalanced data.

Other classification metrics that have been proposed and explored in the literature for data imbalance scenarios are the balanced accuracy and the G-Mean \cite{kubat1997addressing, akosa2017predictive, johnson2019deep, shi2020fault}. Both of these metrics summarize  the recall and specificity, offering a single score that balances the model's capacity to correctly classify observations belonging to both the majority (negative) and the minority (positive) classes. The balanced accuracy and G-mean are measured  as follows:

\begin{equation}
   \text{Balanced Accuracy} = \frac{\text{Recall} + \text{Specificity}}{2}
    \label{eq:6}
\end{equation}

\begin{equation}
   \text{G-Mean} = \sqrt{\text{Recall}\cdot\text{Specificity}}
    \label{eq:7}
\end{equation}

As can be noted from Equations \ref{eq:6} and \ref{eq:7}, both metrics rely solely on the recall and the specificity of a model. The balanced accuracy consists of the arithmetic mean of both metrics, while the G-Mean is their geometric mean. Both metrics can be useful in cases of imbalanced data, as values closer to 1 imply that a model has a high predictive power for both classes.

It can be noted that, while both metrics are similar, due to its mathematical properties, the G-Mean is less sensitive to outliers \cite{akosa2017predictive}. An example can be a model that achieves a perfect specificity of $1$ by correctly classifying all negative samples, but with a low recall of $0.1$. Here, the balanced accuracy would be $0.55$, while the G-Mean would be $0.31$. This shows how the balanced accuracy can be over-optimistic. In this work, the usage of the G-mean as a metric is implemented as it takes into account the rate of true positives and true negatives for malign cases, as its the most under-represented class.

A wide variety of other classification metrics can be used for cases of imbalanced data, like the Matthews correlation coefficient \cite{chicco2020advantages}. This metric corresponds to a correlation coefficient between the observed and predicted classifications. Other metrics  include the Youden's index and the Discriminant Power \cite{sokolova2006beyond}. These metrics, although useful, are not as popular or widely used as the other mentioned classification metrics and might not be as intuitive to understand.

\section{Methods}
\label{methods}

\subsection{Experimental Setup}
\label{exp_setup}

\begin{figure*}
\centering
    \includegraphics[width=1 \textwidth]{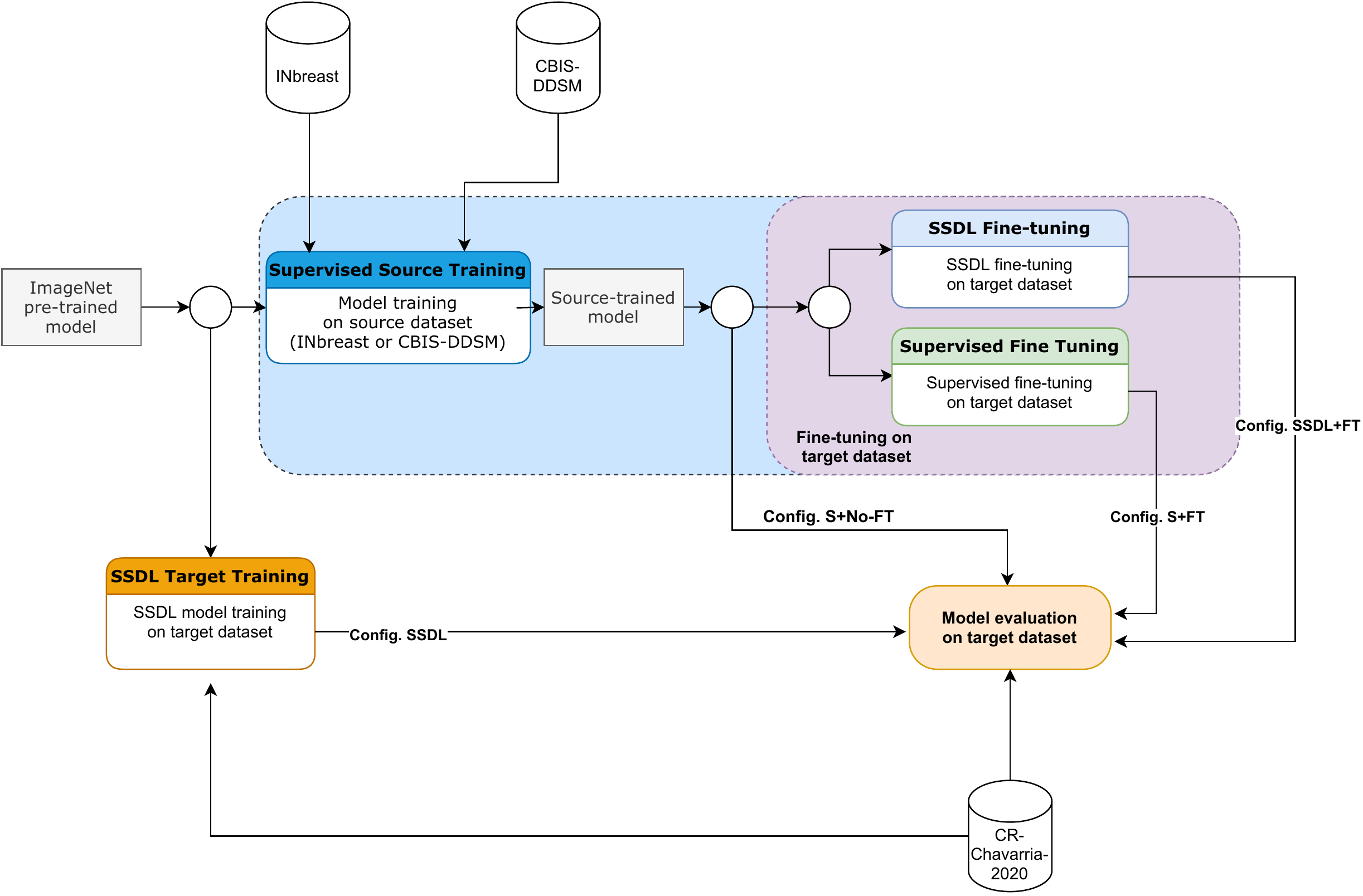}
    \caption{Diagram of experimental configurations presented in this work}
    \label{fig:1}
\end{figure*}

This work proposes the usage of unlabeled data in fine-tuning with the MixMatch \gls{SSDL} approach \cite{berthelot2019mixmatch} . This is done as a mean of improving the performance of deep learning models on the task of binary classification of whole mammogram images under a real-life scenario using a novel target dataset. Evaluations and comparisons are drawn over the performance of deep learning models on the classification of mammogram images obtained in the context of the day-to-day basis of a local medical private clinic of Costa Rica.

For this purpose several experimental configurations were analysed and carried out, as illustrated in Figure \ref{fig:1}. Multiple models were trained under different training configurations to evaluate the impact of \gls{SSDL} on their classification performance on a target dataset. Transfer learning (a simple \enquote{Domain adaptation} method) and loss function based class-imbalance correction were also tested. This was done as means for dealing with common difficulties of the implementation of classification models for real-life use cases, such as limited amounts of data and extreme class imbalance (further detailed in section \ref{DatasetCR}).

Deep learning models were first trained in a supervised manner with complete mammography datasets $D^l_{s,\textrm{IN}}$ and $D^l_{s,\textrm{DDSM}}$ in order to obtain source-trained models, which were further fine-tuned on our target Costarrican dataset in a Supervised (Config. \textbf{S+FT}) or Semi-Supervised (Config. \textbf{SSDL+FT}) manner, with limited amounts of labeled observations $n^l_t$. 

The performance of source-trained models, without fine-tuning on the target dataset, was also evaluated (Config. \textbf{S+No-FT}). The performance of models directly trained on the target dataset using \gls{SSDL}, without domain adaptation from a source mammography dataset (Config. \textbf{SSDL}) was also tested.  Class imbalance correction of the loss function with the \gls{PBC} method developed in \cite{calderon2020correcting} was also used as part of the experiments of Configurations \textbf{SSDL+FT}, \textbf{S+FT} and \textbf{SSDL}. The empirical results obtained in this study showed a considerable impact of its usage for correcting data imbalance. Therefore, we included it to train all of the tested \gls{SSDL} models. Finally, all models were evaluated on test images from our novel target Costarrican dataset.

Due to the extreme data imbalance present in the target dataset ($95\%$ of observations belong to the negative class and $5\%$ to the positive class), specific classification metrics, aside from traditional accuracy, were evaluated as performance indicators. Following the research presented in Section \ref{Metrics}, the G-Mean was chosen as main classification metric. This metric was used to provide insight related to the accuracy of the models on the positive class, without ignoring their predictive power at classifying the negative class. Other metrics including F-2 Score, accuracy, recall, specificity, and precision are also reported.

\gls{DeDiMs} following the novel approach presented by authors in \cite{calderon2021more} were also evaluated, to provide a more thorough analysis of the impact of the choice of source datasets. This method consists in a simple and practical approach to compare different datasets by measuring their dissimilarity in the feature space of a generic deep learning classification model. We aim to quantitatively assess the similarity between the tested datasets and correlate it with the yielded results.

\subsection{Mammography Datasets}
\label{datasets}

\begin{figure}
    \centering
    \begin{tabular}{ccc}
        \begin{subfigure}{0.14\textwidth}
            \includegraphics[width=65pt, height=65pt]{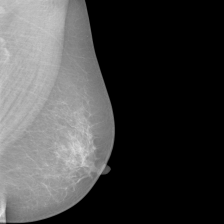}
            \caption{INbreast}
        \end{subfigure}
        &
        \begin{subfigure}{0.14\textwidth}
            \includegraphics[width=65pt, height=65pt]{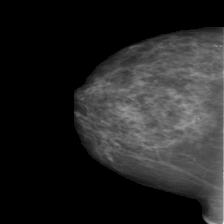}
            \caption{CBIS-DDSM}
        \end{subfigure}
        &
        \begin{subfigure}{0.14\textwidth}
            \includegraphics[width=65pt, height=65pt]{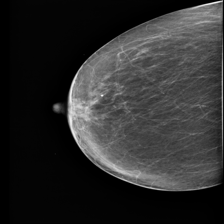}
            
            \caption{CR dataset}
        \end{subfigure} 
        \\

        \begin{subfigure}{0.14\textwidth}
            \includegraphics[width=65pt, height=65pt]{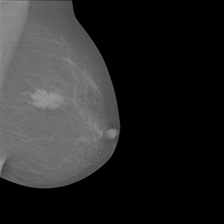}
            \caption{INbreast}
        \end{subfigure}
        &
        \begin{subfigure}{0.14\textwidth}
            \includegraphics[width=65pt, height=65pt]{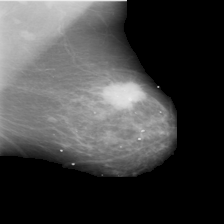}
            \caption{CBIS-DDSM}
        \end{subfigure}
        &
        \begin{subfigure}{0.14\textwidth}
            \includegraphics[width=65pt, height=65pt]{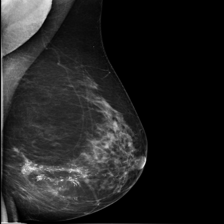}
            \caption{CR dataset}
        \end{subfigure}
        \\
         
    \end{tabular}
    \caption{Examples of benign (top) and malign (bottom) mammogram images from  each dataset}
    \label{fig:9}
\end{figure}

Three different mammography datasets were used to carry out the experiments depicted in this work, summarized in Table \ref{tab:table0}. Sample images are shown  in Figure \ref{fig:9}. The selected datasets correspond to two popular and publicly available \enquote{source} datasets, used solely for model training: the INbreast ($D^l_{s,\textrm{IN}}$) and CBIS-DDSM ($D^l_{s,\textrm{DDSM}}$). A third novel \enquote{target} dataset $D^{l}_{t,\textrm{CR}}$ comprised of mammogram images gathered from a private medical clinic of Costa Rica was also used.

\begin{table}
\caption{Summary of datasets used in this work}
    \label{tab:table0}
\scalebox{0.80}{
\begin{tabular}{lccc}
\hline
 & \textbf{INbreast} \cite{inbreast} & \textbf{CBIS-DDSM} \cite{CBIS-DDSM-citation} & \begin{tabular}[c]{@{}c@{}}\textbf{Target CR}\\ \textbf{Dataset}\end{tabular} \\ \hline
\textbf{Origin} & Portugal & USA & Costa Rica \\[0.5em]   
\textbf{Year} & 2011 & 1997-2016 & 2020 \\[0.5em] 
\multirow{2}{*}{\textbf{\begin{tabular}[c]{@{}l@{}}Number \\ of cases\end{tabular}}} & \multirow{2}{*}{115} & \multirow{2}{*}{1566} & \multirow{2}{*}{87} \\ \\[0.5em] 
\multirow{2}{*}{\textbf{\begin{tabular}[c]{@{}l@{}}Number \\ of images\end{tabular}}} & \multirow{2}{*}{410} & \multirow{2}{*}{3103} & \multirow{2}{*}{341} \\ \\[0.5em] 
\multirow{2}{*}{\textbf{Views}} & CC & CC & CC \\
 & MLO & MLO & MLO \\[0.5em] 
\textbf{Image mode} & \begin{tabular}[c]{@{}c@{}}Full-field\\digital\end{tabular} & \begin{tabular}[c]{@{}c@{}}Digitised\\ screen-film\end{tabular} & \begin{tabular}[c]{@{}c@{}}Full-field\\digital\end{tabular} \\[1em] 
\multirow{3}{*}{\textbf{Categories}} & BI-RADS & BI-RADS & BI-RADS \\
 & ACR Density & ACR Density & \multicolumn{1}{l}{} \\
 &  & Verified Pathology & \multicolumn{1}{l}{} \\ 
\textbf{\begin{tabular}[c]{@{}l@{}}ROI \\ annotations\end{tabular}} & Yes & Yes & No \\ \hline
\end{tabular}}

\end{table}

\subsubsection{Third-party Source Datasets}

Introduced in \cite{inbreast}, the INbreast dataset is a mammographic database comprised of multiple full-field digital mammograms of patients with a wide variety of anomalies like masses and calcifications. Each image is labeled according to the BI-RADS scale from categories 1 to 6 and their density measure with the \gls{ACR} standard. The dataset is composed of 410 images in total, collected from 115 different cases. 

Since this work is focused on the binary classification of mammograms (i.e. according to the presence of breast anomalies), images from the INbreast dataset were divided into 2 groups. Similar to  \cite{shen2019deep}, mammograms labeled with BI-RADS categories 1 and 2 are defined as negative (benign) observations, and the ones labeled with categories 4, 5 and 6 are defined as positive (malign) observations. Mammograms labeled with categories 0 (non-conclusive) and 3 (probably benign) are ignored. For the INbreast dataset, this process results in 287 negative and 100 positive observations.

The \gls{CBIS-DDSM} dataset, presented in \cite{CBIS-DDSM-publication} was made publicly available by the \gls{TCIA}\cite{TCIA-citation}. It corresponds to a curated and standardized version of the DDSM dataset\cite{Heath1998}. The dataset comprises a total of 3103 digitised screen-film mammography images gathered from 1566 cases, labeled according to the type of anomalies present (masses or calcifications), their BI-RADS category, their \gls{ACR} density measure and their verified pathology as benign (1728 images) or malign (1375 images). The dataset presents an overlap between cases that are classified as containing masses or calcifications, as some patients presented both. The total number of images detailed here represents the overall total of both mass and calcification cases, as obtained from \cite{CBIS-DDSM-citation} and subsequently used for model training.

\subsubsection{Cl\'inica Chavarr\'ia's 2020 Mammogram Target Dataset}
\label{DatasetCR}
The \textit{CR-Chavarria-2020} dataset consists of a novel collection of full-field digital mammograms obtained from the Costa Rican medical private clinic Im\'agenes M\'edicas Dr. Chavarr\'ia Estrada, over a period of one year (referred as CR-Chavarria-2020 in Figure \ref{fig:1}). The images are completely anonymized. Specifically, these images correspond to mammograms taken as a result of routinely medical appointments for patients of the clinic across the year 2020. The entire dataset is available for researchers, along with documentation of its distribution, annotations, and extra images that were discarded in the process of constructing the dataset.  If the reader is interested in using our collected dataset, please make contact via email with the first author, as we plan to make the dataset publicly available in the future \footnote{The authors using our novel  dataset are required to cite this paper, for instance as: Calderon-Ramirez, S., Murillo-Hernandez, D., Rojas-Salazar, K., Elizondo, D. A., Yang, S., Moemeni, A., Molina-Cabello, M. (2021). A Real Use Case of Semi-Supervised Learning for Mammogram Classification in a Local Clinic of Costa Rica.}. 

We highlight the value of this dataset as target data for the evaluation of deep learning models in the medical domain, as it is highly representative of the operation conditions that production-implemented models would have to deal with, in a medium sized clinic. The complete dataset, referred as $D^{l}_{t,\textrm{CR}}$, consists of a set of BI-RADS-labeled images. These are also annotated in a similarly manner as the source datasets, with their respective anonymous patient id, gender, age, type of view, and depicted breast. 

The complete $D^{l}_{t,\textrm{CR}}$ dataset contains a total of 341 labeled images from 87 patients. Similarly to the INbreast dataset, images from $D^{l}_{t,\textrm{CR}}$ were also subject to the same \enquote{binarization} process described above. This resulted in the binary-labeled target dataset $D^{b}_{t,\textrm{CR}} \boldsymbol{\subset} D^{l}_{t,\textrm{CR}}$, with a total of 282 images; 268 negative and 14 positive observations from 68 and 4 patients, respectively.

\begin{figure}
\centering
    \includegraphics[width=8.5cm,keepaspectratio]{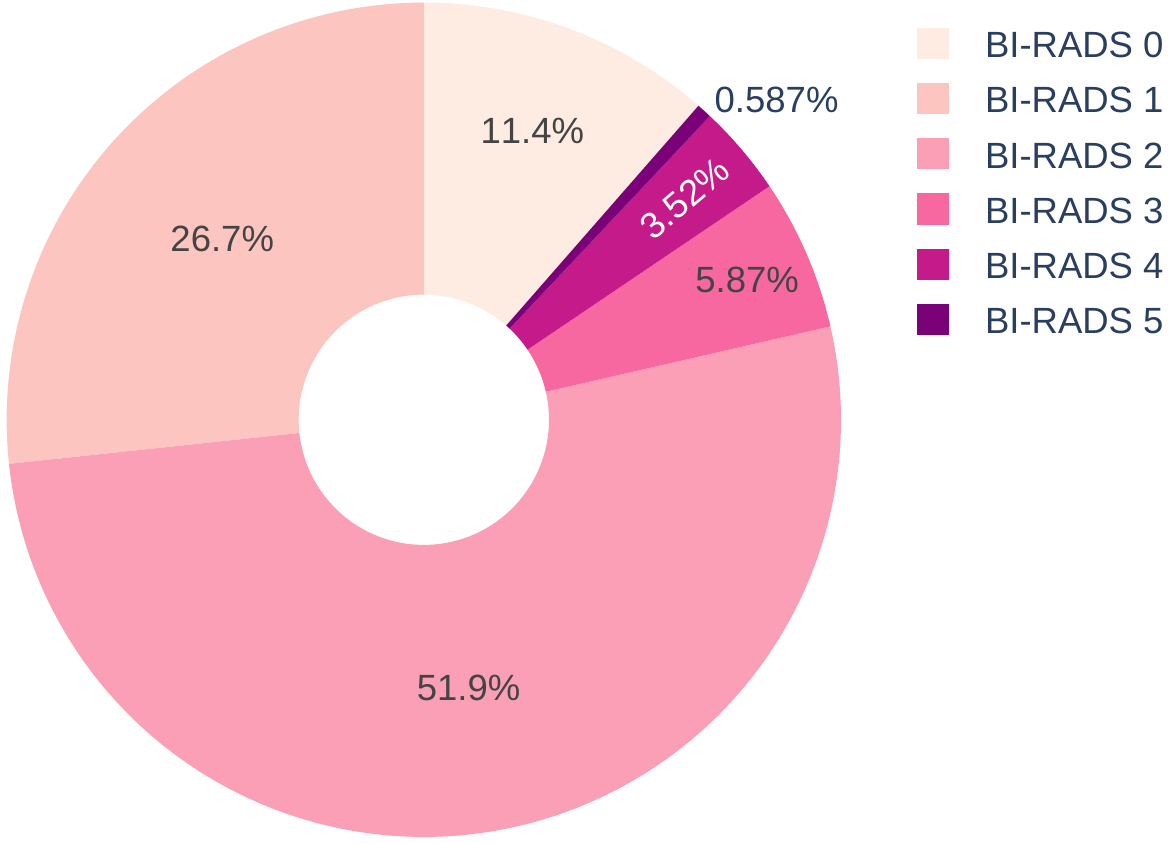}
    \caption{BI-RADS categories distribution for $D^{l}_{t,\textrm{CR}}$}
    \label{fig:2}
\end{figure}

\begin{figure}
\centering
    \includegraphics[width=8.5cm,keepaspectratio]{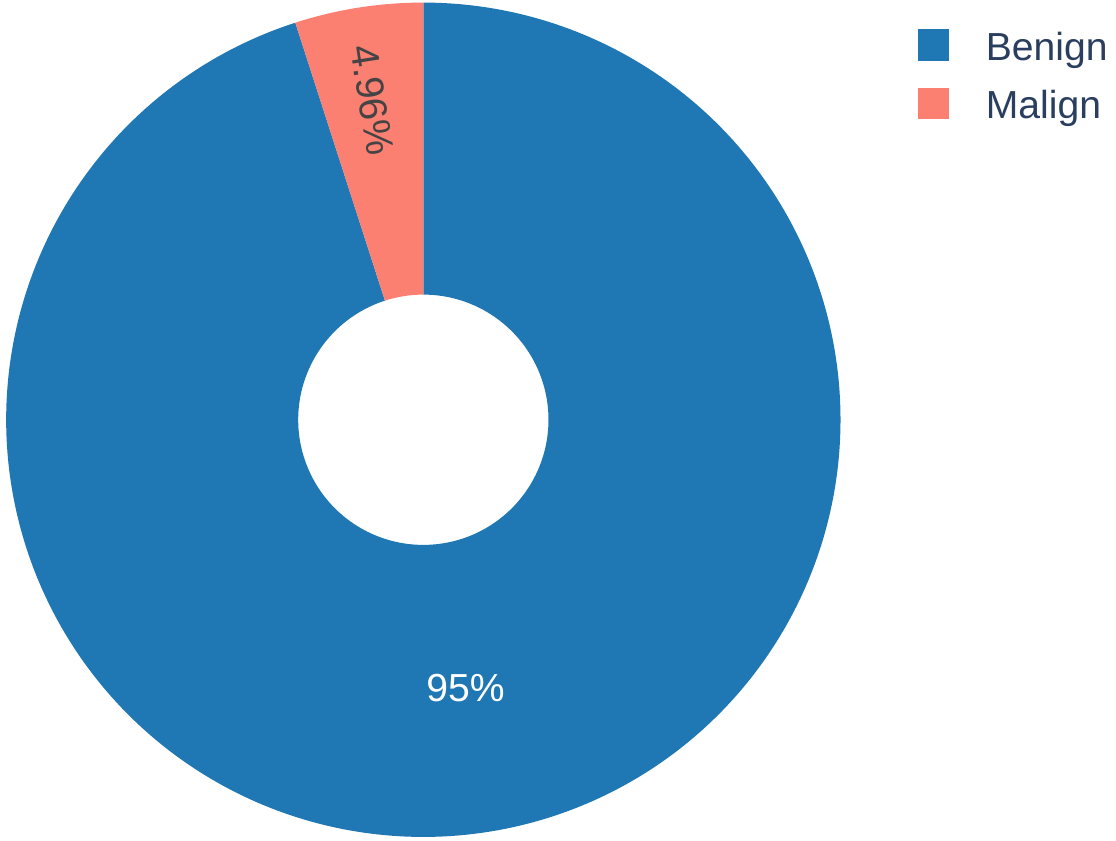}
    \caption{Binary categories distribution for $D^{b}_{t,\textrm{CR}}$}
    \label{fig:3}
\end{figure}

\begin{figure}
\centering
    \includegraphics[width=8.5cm,keepaspectratio]{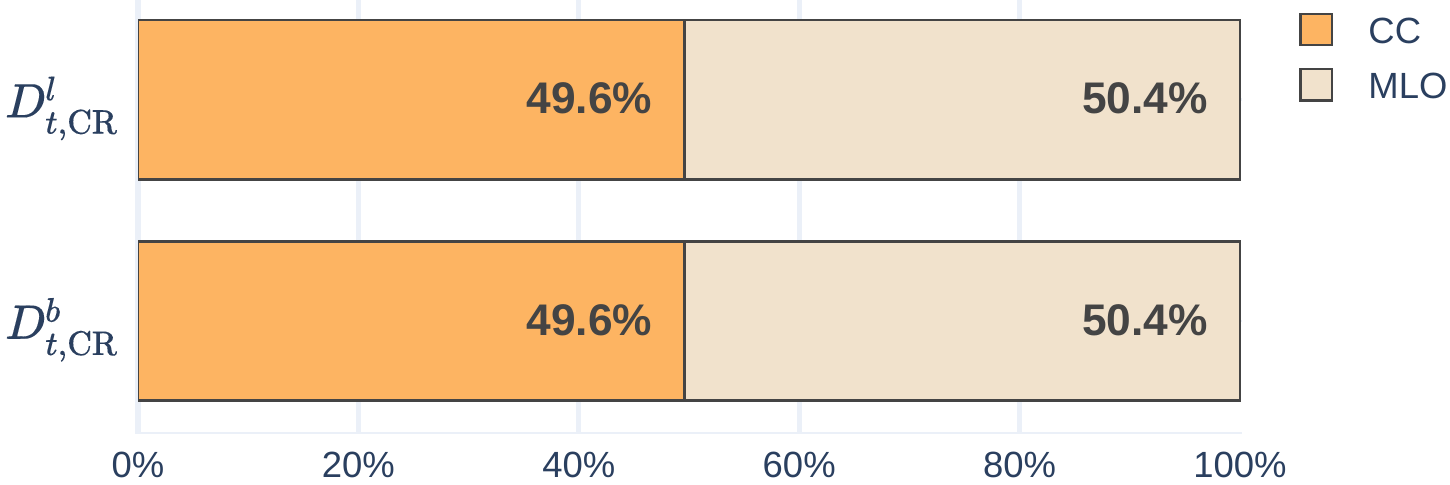}
    \caption{\gls{CC} and \gls{MLO} views distribution for complete and binary-labeled target datasets}
    \label{fig:4}
\end{figure}

\begin{figure}
\centering
    \includegraphics[width=8.5cm,keepaspectratio]{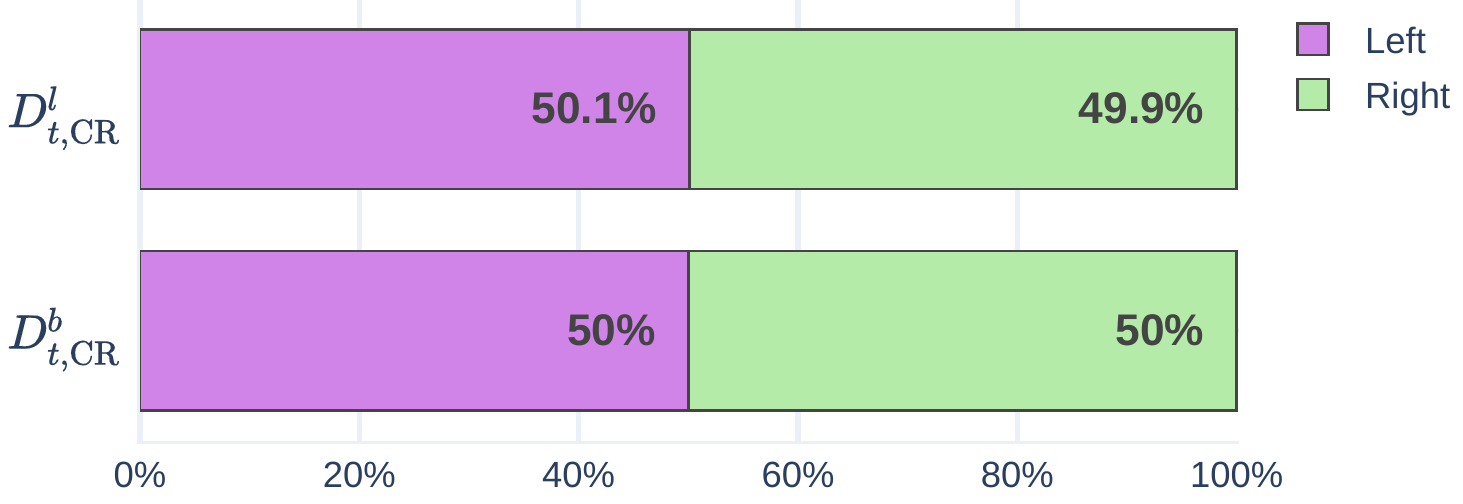}
    \caption{Depicted breast distribution for complete and binary-labeled target datasets}
    \label{fig:5}
\end{figure}

\begin{figure}
\centering
    \includegraphics[width=8.5cm,keepaspectratio]{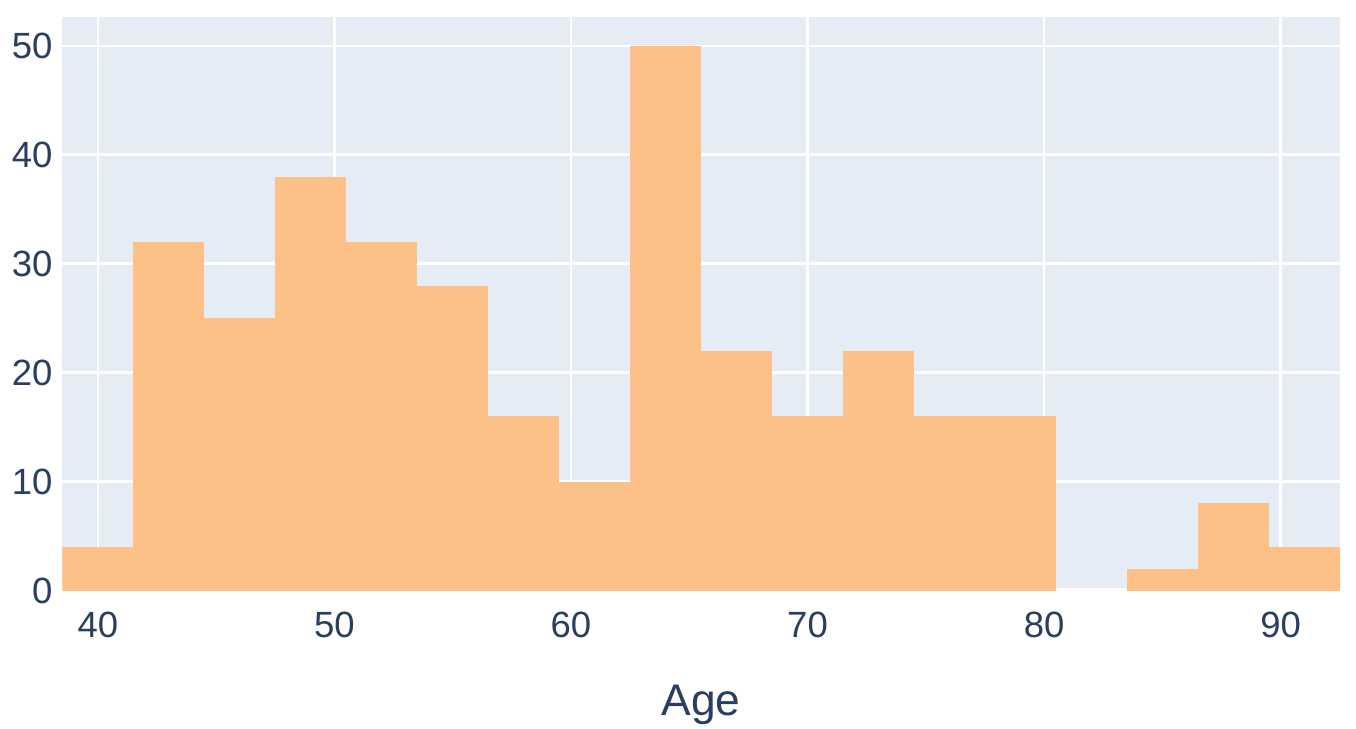}
    \caption{Age distribution for patients in $D^{l}_{t,\textrm{CR}}$}
    \label{fig:6}
\end{figure}

\begin{figure*}
\centering
    \includegraphics[width=12cm,keepaspectratio]{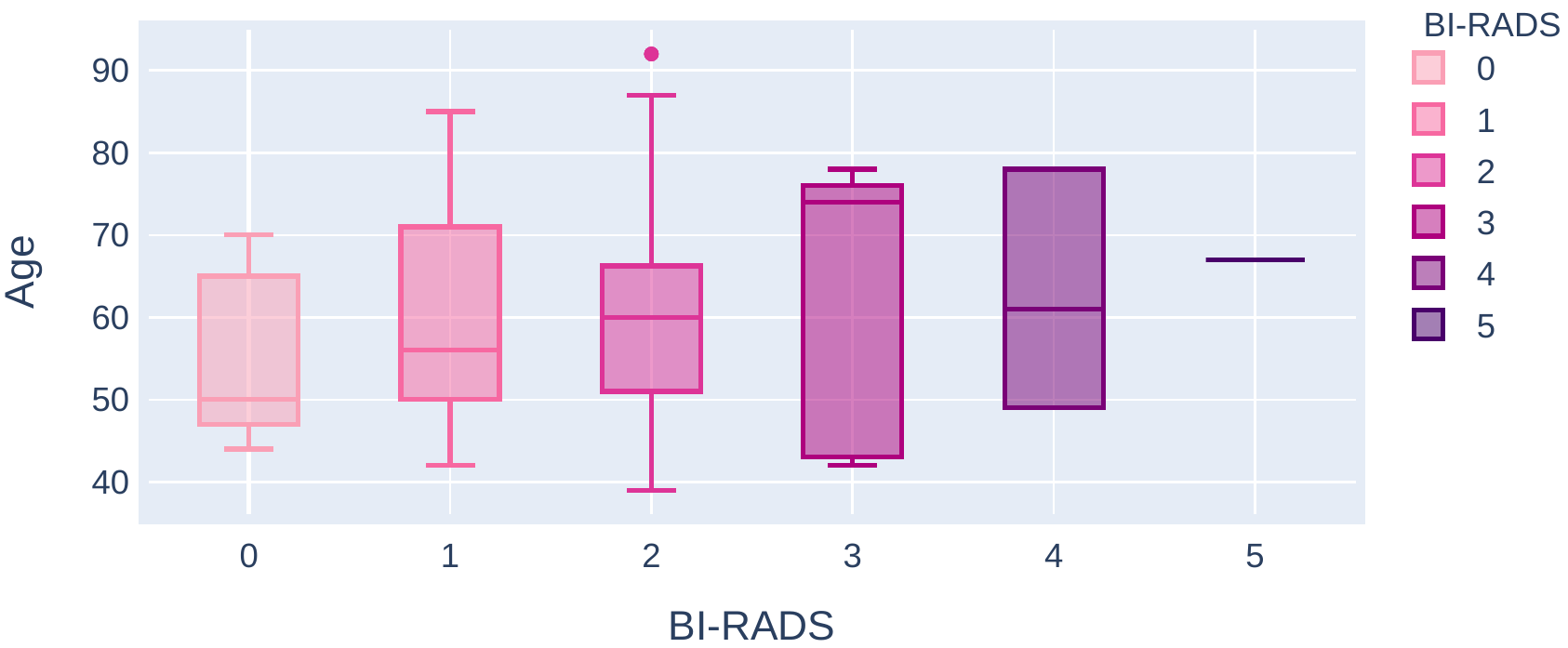}
    \caption{Age distribution according to BI-RADS categories for patients in $D^{l}_{t,\textrm{CR}}$}
    \label{fig:7}
\end{figure*}

\begin{figure*}
\centering
    \includegraphics[width=12cm,keepaspectratio]{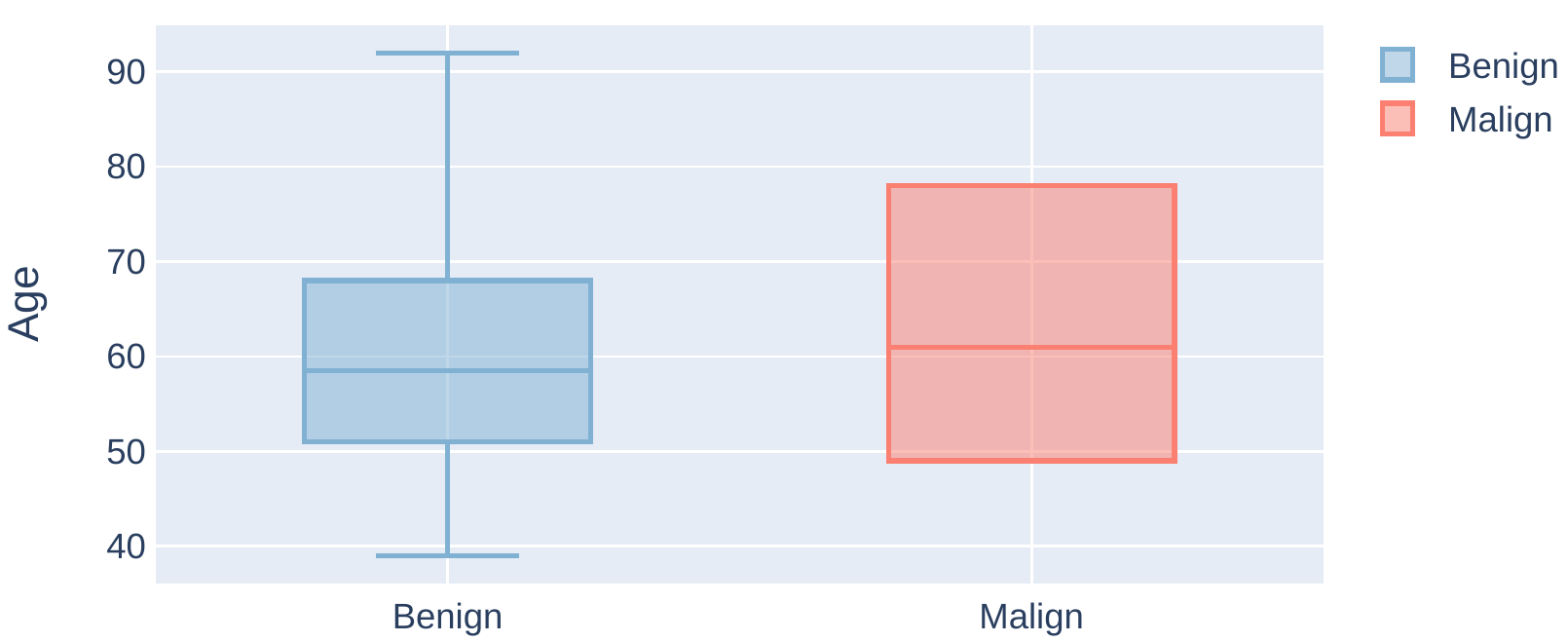}
    \caption{Age distribution according to binary categories for patients in $D^{b}_{t,\textrm{CR}}$}
    \label{fig:8}
\end{figure*}

Figures \ref{fig:2} and \ref{fig:3} illustrate the distribution of both BI-RADS and binary labels for $D^{l}_{t,\textrm{CR}}$ and $D^{b}_{t,\textrm{CR}}$ respectively. Here, the extreme class imbalance of observations can be better appreciated, being one of the most frequent and troublesome situations that arise in the implementation of machine learning models in the medical domain. In addition, figures \ref{fig:4}, \ref{fig:5}, \ref{fig:6}, \ref{fig:7} and \ref{fig:8} show the distribution of other dimensions of both $D^{l}_{t,\textrm{CR}}$ and $D^{b}_{t,\textrm{CR}}$, like the depicted view and breast in each mammogram, along with the age of patients. These aspects show more balanced distributions, as is the case with most mammogram datasets, and that the regular age span for patients varies from 40 to almost 90 years old.

Along with the complete $D^{l}_{t,\textrm{CR}}$ dataset, a set of discarded images has also been made available. These images were  retrieved from the clinic, but were discarded due to low image quality or artifacts (i.e. patients with breast implants). Nevertheless, these could prove to be useful on further investigations, surrounding the robustness of models to domain-specific noise or corruptions in images \cite{hendrycks2019benchmarking}.

\begin{figure}
    \centering
    \begin{tabular}{ccc}
        \begin{subfigure}{0.14\textwidth}
            \includegraphics[width=65pt, height=65pt]{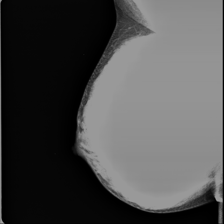}
           
        \end{subfigure}
        
        &
        \begin{subfigure}{0.14\textwidth}
            \includegraphics[width=65pt, height=65pt]{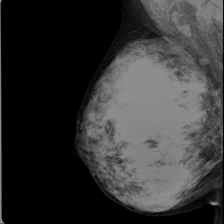}
            
        \end{subfigure}
        &
        \begin{subfigure}{0.14\textwidth}
            \includegraphics[width=65pt, height=65pt]{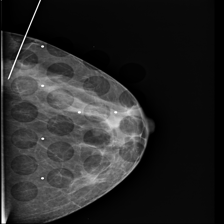}

        \end{subfigure} 
        \vspace{0.25cm}
        \\

        \begin{subfigure}{0.14\textwidth}
            \includegraphics[width=65pt, height=65pt]{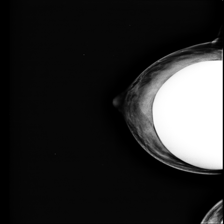}
           
        \end{subfigure}
        &
        \begin{subfigure}{0.14\textwidth}
            \includegraphics[width=65pt, height=65pt]{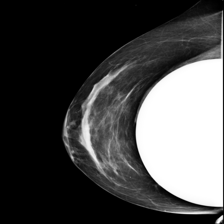}
            
        \end{subfigure}
        &
        \begin{subfigure}{0.14\textwidth}
            \includegraphics[width=65pt, height=65pt]{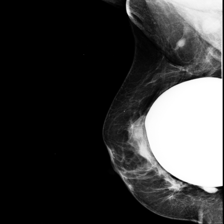}

        \end{subfigure} 
        \\
         
    \end{tabular}
    \caption{Examples of images from original CR data discarded due to image quality (top) or patients with breast implants (bottom)}
    \label{fig:10}
\end{figure}

\subsubsection{Data Preprocessing}

Mammograms from all three described datasets originally possessed considerably high image resolutions. In order to avoid memory constraints, all image files were resized to $224\times224$ pixels, after being converted from the DICOM format to the BMP one. Standardization was applied to all images. The mean and standard deviation, according to the respective dataset employed for training, were calculated (complete INbreast, complete \gls{CBIS-DDSM} or the corresponding training partition of each of the data subsets of the target dataset). Then, for each image, the channel-wise pixel values were subtracted by the mean and divided by the standard deviation. Standardization is done for each training batch. 

\begin{figure}
    \centering
    \begin{tabular}{ccc}
        \begin{subfigure}{0.14\textwidth}
            \includegraphics[width=65pt, height=65pt]{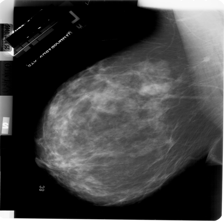}
            \caption{Before}
        \end{subfigure}
        &
        \begin{subfigure}{0.14\textwidth}
            \includegraphics[width=65pt, height=65pt]{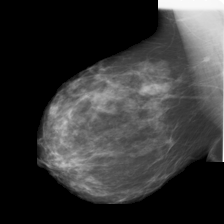}
            \caption{After}
        \end{subfigure}
        
        \\
    
        \begin{subfigure}{0.14\textwidth}
            \includegraphics[width=65pt, height=65pt]{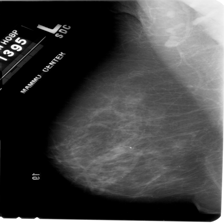}
            \caption{Before}
        \end{subfigure}
        &
        \begin{subfigure}{0.14\textwidth}
            \includegraphics[width=65pt, height=65pt]{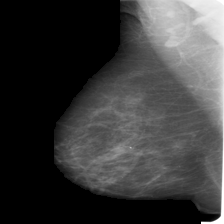}
            \caption{After}
        \end{subfigure}
         
    \end{tabular}
    \caption{Examples of images with background noise from CBIS-DDSM dataset, before and after being preprocessed}
    \label{fig:11}
\end{figure}

Additionally, through visual inspection of the images in \gls{CBIS-DDSM} dataset, it can be noted that several mammograms contain multiple forms of noise, mainly due to the digitization process of the screen-film. Physical labels, orientation tags and scanning artifacts are some of the types of noise inducing elements that can be found in mammogram images, as illustrated in \cite{mustra2016review}.  To minimize the effects of these types of noise, a similar approach to the one described in \cite{beeravolu2021preprocessing} was implemented and applied to images from the \gls{CBIS-DDSM} dataset. This is shown in figure \ref{fig:11}. Authors in \cite{beeravolu2021preprocessing} describe the implemented  preprocessing pipeline in this work, designed for background removal in mammograms. The process consists mainly on the application of a rolling ball algorithm with $radius=5$. This is followed by the application of Huang's fuzzy thresholding and morphological transformations of erosion and dilation. This process results in a binary map that can be used to remove background noise from an image. Such image preprocessing pipeline is implemented in this work, which makes use of the base code made available by the authors of \cite{beeravolu2021preprocessing} and algorithm implementations from the OpenCV library.

\subsubsection{Experiments}

All experiments described in this work were implemented in Python using the FastAI and PyTorch libraries, based on the MixMatch implementation described in \cite{calderon2020dealing}. The PyTorch implementation of the VGG 19-layer with batch normalization was chosen as the main architecture for the models of all experiments. Additionally, experiments of Configurations \textbf{SSDL+FT} and \textbf{S+FT} were also carried out using PyTorch implementations of ResNet-152 and EfficientNet-b0. Transfer learning with pre-trained weights from ImageNet was used for the initial models of all experimental configurations. All depicted experiments were executed employing a total of 10 different randomly generated subsets $D^{b}_{i,t,\textrm{CR}} | i = 1, ..., 10$ of the binary-labeled target Costarrican dataset $D^{b}_{t,\textrm{CR}}$. Each with an average distribution of $70\%$ of images for training and $30\%$ for testing, with observations from different patients for training and for testing. Therefore, around 198 training images (including both labeled and unlabeled), and 82 test images were used. 

The models for the configurations \textbf{SSDL+FT}, \textbf{S+FT} and \textbf{SSDL} were trained on each data subset $D^{b}_{i,t,\textrm{CR}}$, with $n^l_t = 20, 40 \text{ and } 60$ amounts of labeled observations, with $95\%$ of observations corresponding to the negative class (benign) and $5\%$ to the positive class (malign). Class-imbalance correction of the loss function was implemented, respectively, as a weighted cross-entropy loss for the supervised models and as the \gls{PBC} technique \cite{calderon2020correcting} for the \gls{SSDL} models. Supervised models were trained only with the specified $n^l_t$ images from the corresponding training partition of the $D^{b}_{i,t,\textrm{CR}}$ target data subset as $D^l_t$. The \gls{SSDL} models also used the remaining training images in $D^{b}_{i,t,\textrm{CR}}$ as unlabeled data $D^u_t$.

Data augmentation was implemented for the training dataset as random flips and rotations through the FastAI library, for both supervised and \gls{SSDL} models. All models were trained for 50 epochs each, with early stopping to avoid overfitting. We used the G-Mean as a criterion for keeping the model from the epoch with the best score after training. A learning rate of $0.00002$, a weight decay of $0.001$ and a batch size of 10 images were used. The hyper-parameters for MixMatch were set as: $K = 2$ transformations, a sharpening temperature of $T = 0.25$, an alpha mix value of $\alpha = 0.75$ and unsupervised coefficient $\gamma = 200$, following the authors' recommendations in \cite{berthelot2019mixmatch}. The G-Mean, F2-Score, traditional accuracy, recall, specificity, and precision were evaluated for each model, using the test data from their respective $D^{b}_{i,t,\textrm{CR}}$. Results from these metrics were then reported as averages across the 10 target data subsets.

The dissimilarities between the complete source datasets $D^l_{s,\textrm{DDSM}}$ and $D^l_{s,\textrm{IN}}$, and the binary-labeled target dataset $D^{b}_{t,\textrm{CR}}$ were evaluated following the approach presented in \cite{calderon2021more}.  The cosine distance $d_C$ was chosen as the dissimilarity measure, given its reported behavior in \cite{calderon2021more}. This was evaluated in the feature space of a generic Wide-ResNet model pre-trained on ImageNet, with the cosine distance  calculated between the distributions of two datasets on each feature of the feature space and then summed \cite{calderon2021more}. We used 10 randomly selected batches of 40 observations to calculate the feature distribution distances, as suggested in \cite{calderon2021more}.

\section{Results and Discussion}
\label{results}

\begin{table}
\centering
\caption{Classification performance for models of Configuration \textbf{S+No-FT}}
\label{tab:A-3}
\begin{tabular}{lllll}
\hline
\multicolumn{1}{c}{\multirow{2}{*}{Metric}} & \multicolumn{2}{c}{\begin{tabular}[c]{@{}c@{}}INbreast\\ models\end{tabular}} & \multicolumn{2}{c}{\begin{tabular}[c]{@{}c@{}}CBIS-DDSM\\ models\end{tabular}} \\
\multicolumn{1}{c}{} & \multicolumn{1}{c}{$\bar{x}$} & \multicolumn{1}{c}{$s$} & \multicolumn{1}{c}{$\bar{x}$} & \multicolumn{1}{c}{$s$} \\ \hline
G-Mean & 0.3773 & 0.1043 & 0.3476 & 0.2534 \\
F2-Score & 0.1882 & 0.0625 & 0.1347 & 0.1148 \\
Accuracy & 0.2183 & 0.0602 & 0.7379 & 0.0678 \\
Recall & 0.7667 & 0.2509 & 0.2333 & 0.1876 \\
Specificity & 0.1901 & 0.0558 & 0.7639 & 0.0707 \\
Precision & 0.0470 & 0.0160 & 0.0517 & 0.0467 \\ \hline
\end{tabular}

\end{table}

\begin{table}
\centering
\caption{Classification performance for models of Configuration \textbf{SSDL}}
\label{tab:B}

\scalebox{0.85}{
\begin{tabular}{lllllll}
\hline
\multicolumn{1}{c}{\multirow{2}{*}{Metric}} & \multicolumn{2}{c}{$n^l_t = 20$} & \multicolumn{2}{c}{$n^l_t = 40$} & \multicolumn{2}{c}{$n^l_t = 60$} \\
\multicolumn{1}{c}{} & \multicolumn{1}{c}{$\bar{x}$} & \multicolumn{1}{c}{$s$} & \multicolumn{1}{c}{$\bar{x}$} & \multicolumn{1}{c}{$s$} & \multicolumn{1}{c}{$\bar{x}$} & \multicolumn{1}{c}{$s$} \\ \hline
G-Mean & 0.4798 & 0.1936 & 0.5720 & 0.1257 & 0.6413 & 0.0929 \\
F2-Score & 0.2169 & 0.1194 & 0.2683 & 0.1168 & 0.3038 & 0.0889 \\
Accuracy & 0.5786 & 0.2212 & 0.6482 & 0.2172 & 0.6869 & 0.1412 \\
Recall & 0.5167 & 0.2687 & 0.5750 & 0.2648 & 0.6333 & 0.2297 \\
Specificity & 0.5815 & 0.2404 & 0.6518 & 0.2354 & 0.6904 & 0.1544 \\
Precision & 0.1189 & 0.1551 & 0.1079 & 0.0754 & 0.1096 & 0.0491 \\ \hline
\end{tabular}}

\end{table}

The results of each of the described experimental configurations are presented in Tables \ref{tab:summary_A-1_A-2}, \ref{tab:A-3}, \ref{tab:B}, \ref{tab:A_INbreast}, and \ref{tab:A_DDSM}, as the mean and standard deviation of the corresponding classification metrics, evaluated across each of the 10 random data subsets of the target dataset. Results are also presented accordingly to the number of $n^l_t$ that were used for training (Configs. \textbf{SSDL+FT}, \textbf{S+FT} and \textbf{SSDL}).

The  classification performance on the target dataset of source-trained-only models appears to be rather poor, with no clear advantages between the source datasets, as seen in Table \ref{tab:A-3}. The low average G-Mean values yielded by models trained on each of the source datasets show a deficient ability to correctly discriminate between both classes. This situation is confirmed by the yielded average recall and specificity values, which show a clear imbalance of the discrimination accuracy for each class. Low average F2-Score values also reinforce this conclusion, showing a relatively high number of \gls{FP} in proportion to \gls{TP} predictions.  The \enquote{accuracy paradox} can also be seen in the yielded average accuracy scores of Table \ref{tab:A-3}. Models trained on $D^l_{s,\textrm{IN}}$ scored notably lower accuracy values in comparison to models trained on $D^l_{s,\textrm{DDSM}}$. However, further analysis suggests that the higher accuracy scores of the latter models were due to their relatively high specificity scores. This shows a clear bias on the accuracy scores for the majority class (negative cases).

Table \ref{tab:B} shows the classification performance results of models trained with \gls{SSDL} on the target dataset, without domain adaptation from a source mammography dataset. Considerably high standard deviations are observed for the majority of the results. Despite this, the average values of both G-Mean and F2-Score show steady improvements as the number of $n^l_t$ increases. It is only logical that these models are able to make a better use of an increased number of labeled observations for training. This is mainly due to the fact that they do not possess previous domain-knowledge from a source dataset.

Significant improvements can be perceived in the classification performance of the source-trained models after fine-tuning on the target dataset, as depicted by Tables \ref{tab:A_INbreast} and \ref{tab:A_DDSM}. Wilcoxon signed-rank tests were applied to these results in order to identify statistically significant ($p \text{-values} < 0.1$) differences between the performance of the models fine-tuned either in a supervised manner or with the \gls{SSDL} method. Table \ref{tab:A_INbreast} shows the results of the models first trained on $D^l_{s,\textrm{IN}}$ and then fine-tuned on the target dataset. It includes all the tested model architectures. Models fine-tuned with \gls{SSDL} generally yielded moderately better average G-Mean and F2-Score results in comparison to models fine-tuned using a supervised manner. This happens specially when using a reduced number of labeled observations for training ($n^l_t = 20,40$), as the perceived gains decrease with a higher value of $n^l_t$. With more labels,  the  results tend to reveal less statistical significance with $p \text{-values} > 0.1$.

The results shown in Table \ref{tab:A_DDSM} correspond to the models that were first trained on $D^l_{s,\textrm{DDSM}}$ and then fine-tuned on the target dataset. Considerably higher average G-Mean and F2-Score values were yielded by models fine-tuned with \gls{SSDL}. They show statistical significance when employing lower amounts of labeled observations ($n^l_t = 20, 40$), specially for the models that used the VGG19 architecture. For these models, the ones that were fine-tuned in a supervised fashion scored higher average specificity values. However, by observing their respective average recall values it is clear that their rate of correct predictions is unbalanced for both classes. These models appear to be biased to the majority class. However, the models with \gls{SSDL} can be considered to be less biased, according to the yielded results. Their average recall and specificity show a more stable behavior. A similar, yet not so strong trend can be noticed in models of the other tested architectures. Models with supervised fine-tuning also achieved generally higher average accuracy values, when compared to the no fine-tuned models.

In summary, models that were subject to domain adaptation from a source mammography dataset showed improved classification performance results in comparison to the other experimental configurations tested in this work. However, the choice of source dataset and deep learning model architecture are shown to be important factors in the yielded results.  Models that used the \gls{CBIS-DDSM} as source dataset showed better overall results, with more evident trends and noticeable improvements by the use of \gls{SSDL}. Models that used the INbreast as source dataset scored relatively worse results, with no significant differences between the performance of supervised and \gls{SSDL} models. Additionally, the performance of supervised models does not  change significantly across the different number of  labeled observations tested. These models achieved seemingly converging G-Mean values with fairly balanced recall and specificity values from a lower number of $n^l_t$. This was observed on all tested model architectures.

Regarding the poor performance of configuration \textbf{S+No-FT}, we found that the measurement of the  \gls{DeDiMs} can be an useful warning of choosing one unlabeled data source over another. The dissimilarity between $D^l_{s,\textrm{IN}}$ and $D^{b}_{t,\textrm{CR}}$ was measured as $\boldsymbol{31.10}\pm 1.56$, while for the dissimilarity between $D^l_{s,\textrm{DDSM}}$ and $D^{b}_{t,\textrm{CR}}$ was $\boldsymbol{26.21}\pm 2.31$, both results with $p \text{-values} < 0.05$. These results indicate that the feature distributions (using a generic ImageNet pre-trained model) between both source datasets and the target dataset are significantly different. This can explain the poor results of Configuration \textbf{S+No-FT} as a high dissimilarity is accurately suggesting that some sort of domain adaption is needed. At the same time, a lower dissimilarity between $D^l_{s,\textrm{DDSM}}$ and $D^{b}_{t,\textrm{CR}}$ might indicate that the former could be better suited to be used as a source dataset, as seen in the yielded performance behavior for both datasets  in  Tables \ref{tab:A_INbreast} and \ref{tab:A_DDSM}. The reasons behind a higher dissimilarity between two datasets need to be explored further.

Table \ref{tab:summary_A-1_A-2} summarizes the performance of the models with the lowest number of labels. The average G-Mean scores are shown for models fine-tuned with the lowest number of labeled observations. The results in Table \ref{tab:summary_A-1_A-2}  show how the model architecture constitutes an important factor in the yielded performance of the models. As seen previously,  \gls{SSDL} models show better performance in comparison to supervised ones. However, the improved gains are stronger for the more complex models (i.e. architectures with more trainable parameters).

Overall, \gls{SSDL} models without domain adaptation show significantly lower performance than models with domain adaptation either supervised or with \gls{SSDL} (Configs. \textbf{S+FT} and \textbf{SSDL+FT}). Low average precision and F2-Score values are observed for models of all experimental configurations. As it was mentioned, for a binary classification task, this implies a considerably high number of false positives in relation to the number of true positives. Nonetheless, it must be taken into account that the target dataset suffers from extreme class imbalance. This causes the calculation of the precision to be highly sensitive to the number of false positives. 


\renewcommand{\thefootnote}{\fnsymbol{footnote}}

\begin{table}[H]

\centering
\caption[]{Results of Configurations \textbf{SSDL+FT} and \textbf{S+FT}, using \textbf{INbreast} as source dataset}
\label{tab:A_INbreast}

\subfloat[VGG19 with batch normalization]{
\scalebox{0.9}{
\begin{tabular}{clrrrr}
\hline
\multicolumn{1}{c}{\multirow{2}{*}{$n^l_t$}} & \multicolumn{1}{c}{\multirow{2}{*}{Metric}} & \multicolumn{2}{c}{SSDL} & \multicolumn{2}{c}{Supervised} \\ 
\multicolumn{1}{c}{} & \multicolumn{1}{c}{} & \multicolumn{1}{c}{$\bar{x}$} & \multicolumn{1}{c}{$s$} & \multicolumn{1}{c}{$\bar{x}$} & \multicolumn{1}{c}{$s$} \\ \hline
\multirow{6}{*}{20} & G-Mean & \textbf{0.6764} & 0.1084 & 0.6682 & 0.0770 \\
 & F2-Score & \textbf{0.3506} & 0.0973 & 0.3133 & 0.0673 \\
 & Accuracy\footnotemark[1] & \textbf{0.7812} & 0.0727 & 0.7014 & 0.0793 \\
 & Recall & 0.5917 & 0.1687 & \textbf{0.6500} & 0.1748 \\
 & Specificity\footnotemark[1] & \textbf{0.7907} & 0.0755 & 0.7048 & 0.0876 \\
 & Precision\footnotemark[1] & \textbf{0.1436} & 0.0636 & 0.1074 & 0.0335 \\ \hline
\multirow{6}{*}{40} & G-Mean & \textbf{0.7017} & 0.0932 & 0.6656 & 0.0877 \\
 & F2-Score & \textbf{0.3650} & 0.0899 & 0.3484 & 0.1112 \\
 & Accuracy & \textbf{0.7742} & 0.0659 & 0.7224 & 0.1590 \\
 & Recall & \textbf{0.6417} & 0.1715 & 0.6417 & 0.2081 \\
 & Specificity & \textbf{0.7810} & 0.0693 & 0.7262 & 0.1721 \\
 & Precision & 0.1380 & 0.0373 & \textbf{0.1837} & 0.1708 \\ \hline
\multirow{6}{*}{60} & G-Mean & \textbf{0.6689} & 0.0957 & 0.6604 & 0.0876 \\
 & F2-Score & 0.3278 & 0.0958 & \textbf{0.3415} & 0.1116 \\
 & Accuracy & 0.7211 & 0.1169 & \textbf{0.7432} & 0.1374 \\
 & Recall & \textbf{0.6250} & 0.1318 & 0.6000 & 0.1748 \\
 & Specificity & 0.7267 & 0.1230 & \textbf{0.7510} & 0.1466 \\
 & Precision & 0.1226 & 0.0565 & \textbf{0.1822} & 0.1704 \\ \hline
\end{tabular}}
}

\subfloat[ResNet-152]{
\scalebox{0.9}{
\begin{tabular}{clrrrr}
\hline
\multicolumn{1}{c}{\multirow{2}{*}{$n^l_t$}} & \multicolumn{1}{c}{\multirow{2}{*}{Metric}} & \multicolumn{2}{c}{SSDL} & \multicolumn{2}{c}{Supervised} \\
\multicolumn{1}{c}{} & \multicolumn{1}{c}{} & \multicolumn{1}{c}{$\bar{x}$} & \multicolumn{1}{c}{$s$} & \multicolumn{1}{c}{$\bar{x}$} & \multicolumn{1}{c}{$s$} \\ \hline
\multirow{6}{*}{20} & G-Mean & \textbf{0.6774} & 0.1167 & 0.6767 & 0.1021 \\
 & F2-Score & \textbf{0.3577} & 0.1349 & 0.3155 & 0.0828 \\
 & Accuracy & \textbf{0.6851} & 0.1903 & 0.5978 & 0.1021 \\
 & Recall & 0.7250 & 0.2189 & \textbf{0.7917} & 0.1632 \\
 & Specificity & \textbf{0.6838} & 0.2064 & 0.5883 & 0.1062 \\
 & Precision\footnotemark[1] & \textbf{0.1407} & 0.0869 & 0.0946 & 0.0300 \\ \hline
\multirow{6}{*}{40} & G-Mean & \textbf{0.6762} & 0.1170 & 0.6705 & 0.1037 \\
 & F2-Score & \textbf{0.3422} & 0.1285 & 0.3099 & 0.0814 \\
 & Accuracy & \textbf{0.7075} & 0.1237 & 0.6443 & 0.0940 \\
 & Recall & 0.6750 & 0.2058 & \textbf{0.7250} & 0.2189 \\
 & Specificity & \textbf{0.7080} & 0.1341 & 0.6409 & 0.1017 \\
 & Precision & \textbf{0.1230} & 0.0640 & 0.0975 & 0.0283 \\ \hline
\multirow{6}{*}{60} & G-Mean & 0.6539 & 0.1384 & \textbf{0.6774} & 0.0787 \\
 & F2-Score & 0.3226 & 0.1386 & \textbf{0.3388} & 0.0973 \\
 & Accuracy & 0.6590 & 0.1547 & \textbf{0.7034} & 0.1353 \\
 & Recall & 0.6667 & 0.2041 & \textbf{0.6667} & 0.1179 \\
 & Specificity & 0.6584 & 0.1599 & \textbf{0.7051} & 0.1442 \\
 & Precision & 0.1201 & 0.0803 & \textbf{0.1231} & 0.0567 \\ \hline
\end{tabular}}
}

\subfloat[EfficientNet-b0]{
\scalebox{0.9}{
\begin{tabular}{clrrrr}
\hline
\multicolumn{1}{c}{\multirow{2}{*}{$n^l_t$}} & \multicolumn{1}{c}{\multirow{2}{*}{Metric}} & \multicolumn{2}{c}{SSDL} & \multicolumn{2}{c}{Supervised} \\
\multicolumn{1}{c}{} & \multicolumn{1}{c}{} & \multicolumn{1}{c}{$\bar{x}$} & \multicolumn{1}{c}{$s$} & \multicolumn{1}{c}{$\bar{x}$} & \multicolumn{1}{c}{$s$} \\ \hline
\multirow{6}{*}{20} & G-Mean & \textbf{0.6512} & 0.1081 & 0.6393 & 0.0603 \\
 & F2-Score & \textbf{0.3282} & 0.1142 & 0.2954 & 0.0538 \\
 & Accuracy & \textbf{0.7440} & 0.1433 & 0.7011 & 0.1072 \\
 & Recall & \textbf{0.6000} & 0.2108 & 0.5917 & 0.1208 \\
 & Specificity & \textbf{0.7519} & 0.1565 & 0.7062 & 0.1157 \\
 & Precision & \textbf{0.1309} & 0.0624 & 0.1041 & 0.0338 \\ \hline
\multirow{6}{*}{40} & G-Mean\footnotemark[1] & \textbf{0.6752} & 0.0977 & 0.6401 & 0.0754 \\
 & F2-Score & \textbf{0.3331} & 0.0999 & 0.2920 & 0.0598 \\
 & Accuracy & 0.7090 & 0.1174 & \textbf{0.7139} & 0.0768 \\
 & Recall & \textbf{0.6667} & 0.2041 & 0.5750 & 0.1208 \\
 & Specificity & 0.7107 & 0.1273 & \textbf{0.7209} & 0.0797 \\
 & Precision & \textbf{0.1161} & 0.0470 & 0.0999 & 0.0232 \\ \hline
\multirow{6}{*}{60} & G-Mean\footnotemark[1] & \textbf{0.6620} & 0.0919 & 0.6422 & 0.0869 \\
 & F2-Score & \textbf{0.3274} & 0.1031 & 0.2975 & 0.0808 \\
 & Accuracy & \textbf{0.7229} & 0.1279 & 0.7161 & 0.0836 \\
 & Recall & \textbf{0.6250} & 0.1768 & 0.5750 & 0.1208 \\
 & Specificity & \textbf{0.7282} & 0.1376 & 0.7234 & 0.0851 \\
 & Precision & \textbf{0.1196} & 0.0540 & 0.1036 & 0.0352 \\ \hline
\end{tabular}}
}
\end{table}
\footnotetext[1]{Statistic significance ($p \text{-values} < 0.1$) for average differences between results of \gls{SSDL} and supervised models}

\begin{table}[H]
\centering
\caption[]{Results of Configurations \textbf{SSDL+FT} and \textbf{S+FT}, using \textbf{\gls{CBIS-DDSM}} as source dataset}
\label{tab:A_DDSM}

\subfloat[VGG19 with batch normalization]{
\scalebox{0.9}{
\begin{tabular}{clrrrr}
\hline
\multicolumn{1}{c}{\multirow{2}{*}{$n^l_t$}} & \multicolumn{1}{c}{\multirow{2}{*}{Metric}} & \multicolumn{2}{c}{SSDL} & \multicolumn{2}{c}{Supervised} \\  
\multicolumn{1}{c}{} & \multicolumn{1}{c}{} & \multicolumn{1}{c}{$\bar{x}$} & \multicolumn{1}{c}{$s$} & \multicolumn{1}{c}{$\bar{x}$} & \multicolumn{1}{c}{$s$} \\ \hline
\multirow{6}{*}{20} & G-Mean\footnotemark[1] & \textbf{0.7313} & 0.0742 & 0.5163 & 0.2826 \\
 & F2-Score\footnotemark[1] & \textbf{0.3910} & 0.0909 & 0.2892 & 0.1797 \\
 & Accuracy\footnotemark[1] & 0.7455 & 0.1115 & \textbf{0.8333} & 0.0710 \\
 & Recall\footnotemark[1] & \textbf{0.7333} & 0.1459 & 0.3917 & 0.2292 \\
 & Specificity\footnotemark[1] & 0.7460 & 0.1201 & \textbf{0.8554} & 0.0709 \\
 & Precision & 0.1480 & 0.0551 & \textbf{0.1602} & 0.1289 \\ \hline
\multirow{6}{*}{40} & G-Mean\footnotemark[1] & \textbf{0.7264} & 0.0909 & 0.5743 & 0.2308 \\
 & F2-Score\footnotemark[1] & \textbf{0.3917} & 0.1124 & 0.3070 & 0.1597 \\
 & Accuracy\footnotemark[1] & 0.7588 & 0.1041 & \textbf{0.8286} & 0.0476 \\
 & Recall\footnotemark[1] & \textbf{0.7083} & 0.1632 & 0.4417 & 0.2189 \\
 & Specificity\footnotemark[1] & 0.7612 & 0.1110 & \textbf{0.8482} & 0.0453 \\
 & Precision & \textbf{0.1520} & 0.0630 & 0.1458 & 0.0899 \\ \hline
\multirow{6}{*}{60} & G-Mean & \textbf{0.7142} & 0.0717 & 0.6466 & 0.1462 \\
 & F2-Score & \textbf{0.3779} & 0.1001 & 0.3436 & 0.1506 \\
 & Accuracy\footnotemark[1] & 0.7197 & 0.1445 & \textbf{0.8132} & 0.0723 \\
 & Recall\footnotemark[1] & \textbf{0.7333} & 0.1459 & 0.5333 & 0.2297 \\
 & Specificity\footnotemark[1] & 0.7190 & 0.1559 & \textbf{0.8271} & 0.0779 \\
 & Precision & 0.1435 & 0.0623 & \textbf{0.1539} & 0.0834 \\ \hline
\end{tabular}}
}

\subfloat[ResNet-152]{
\scalebox{0.9}{
\begin{tabular}{clrrrr}
\hline
\multicolumn{1}{c}{\multirow{2}{*}{$n^l_t$}} & \multicolumn{1}{c}{\multirow{2}{*}{Metric}} & \multicolumn{2}{c}{SSDL} & \multicolumn{2}{c}{Supervised} \\  
\multicolumn{1}{c}{} & \multicolumn{1}{c}{} & \multicolumn{1}{c}{$\bar{x}$} & \multicolumn{1}{c}{$s$} & \multicolumn{1}{c}{$\bar{x}$} & \multicolumn{1}{c}{$s$} \\ \hline
\multirow{6}{*}{20} & G-Mean\footnotemark[1] & \textbf{0.6575} & 0.1075 & 0.5857 & 0.0598 \\
 & F2-Score & \textbf{0.3068} & 0.1151 & 0.2527 & 0.0672 \\
 & Accuracy\footnotemark[1] & 0.5910 & 0.1184 & \textbf{0.6828} & 0.1286 \\
 & Recall\footnotemark[1] & \textbf{0.7583} & 0.1687 & 0.5000 & 0.0000 \\
 & Specificity\footnotemark[1] & 0.5825 & 0.1234 & \textbf{0.6925} & 0.1354 \\
 & Precision & \textbf{0.0959} & 0.0538 & 0.0917 & 0.0432 \\ \hline
\multirow{6}{*}{40} & G-Mean & \textbf{0.5947} & 0.1502 & 0.5759 & 0.0626 \\
 & F2-Score & \textbf{0.2610} & 0.1228 & 0.2426 & 0.0643 \\
 & Accuracy & 0.6541 & 0.0909 & \textbf{0.6616} & 0.1327 \\
 & Recall & \textbf{0.5833} & 0.2966 & 0.5000 & 0.0000 \\
 & Specificity & 0.6569 & 0.1002 & \textbf{0.6703} & 0.1397 \\
 & Precision & 0.0839 & 0.0416 & \textbf{0.0848} & 0.0368 \\ \hline
\multirow{6}{*}{60} & G-Mean & \textbf{0.6153} & 0.1250 & 0.5561 & 0.0769 \\
 & F2-Score & \textbf{0.2706} & 0.1003 & 0.2207 & 0.0642 \\
 & Accuracy & \textbf{0.6869} & 0.0613 & 0.6463 & 0.0909 \\
 & Recall & \textbf{0.5750} & 0.2372 & 0.4750 & 0.0791 \\
 & Specificity & \textbf{0.6931} & 0.0696 & 0.6554 & 0.0940 \\
 & Precision & \textbf{0.0907} & 0.0399 & 0.0731 & 0.0306 \\ \hline
\end{tabular}}
}

\subfloat[EfficientNet-b0]{
\scalebox{0.9}{
\begin{tabular}{clrrrr}
\hline
\multicolumn{1}{c}{\multirow{2}{*}{$n^l_t$}} & \multicolumn{1}{c}{\multirow{2}{*}{Metric}} & \multicolumn{2}{c}{SSDL} & \multicolumn{2}{c}{Supervised} \\
\multicolumn{1}{c}{} & \multicolumn{1}{c}{} & \multicolumn{1}{c}{$\bar{x}$} & \multicolumn{1}{c}{$s$} & \multicolumn{1}{c}{$\bar{x}$} & \multicolumn{1}{c}{$s$} \\ \hline
\multirow{6}{*}{20} & G-Mean & \textbf{0.5982} & 0.0753 & 0.5824 & 0.0489 \\
 & F2-Score & \textbf{0.2521} & 0.0550 & 0.2467 & 0.0475 \\
 & Accuracy & \textbf{0.6537} & 0.0938 & 0.5886 & 0.1426 \\
 & Recall & 0.5750 & 0.2058 & \textbf{0.6083} & 0.1419 \\
 & Specificity & \textbf{0.6579} & 0.1060 & 0.5866 & 0.1558 \\
 & Precision & \textbf{0.0805} & 0.0181 & 0.0757 & 0.0192 \\ \hline
\multirow{6}{*}{40} & G-Mean & 0.6055 & 0.1048 & \textbf{0.6130} & 0.0519 \\
 & F2-Score & 0.2607 & 0.0853 & \textbf{0.2884} & 0.0668 \\
 & Accuracy & 0.6693 & 0.0894 & \textbf{0.7182} & 0.1443 \\
 & Recall & \textbf{0.5750} & 0.2372 & 0.5333 & 0.1054 \\
 & Specificity & 0.6742 & 0.0993 & \textbf{0.7264} & 0.1562 \\
 & Precision & 0.0839 & 0.0262 & \textbf{0.1102} & 0.0436 \\ \hline
\multirow{6}{*}{60} & G-Mean & \textbf{0.6014} & 0.0971 & 0.6013 & 0.0618 \\
 & F2-Score & 0.2612 & 0.0835 & \textbf{0.2792} & 0.0882 \\
 & Accuracy & 0.6810 & 0.1020 & \textbf{0.7187} & 0.1391 \\
 & Recall & \textbf{0.5500} & 0.1972 & 0.5000 & 0.0000 \\
 & Specificity & 0.6878 & 0.1106 & \textbf{0.7300} & 0.1459 \\
 & Precision & 0.0890 & 0.0362 & \textbf{0.1200} & 0.0869 \\ \hline
\end{tabular}}
}

\end{table}


\begin{table*}
\centering
\caption{Summary of G-Mean scores for models of Configs. \textbf{SSDL+FT} and \textbf{S+FT}, using $n^l_t = 20$ labeled observations. The corresponding number of trainable parameters for the PyTorch-implementation of each architecture is also shown}
\label{tab:summary_A-1_A-2}

\begin{tabular}{lllllllllc}
\hline
\multicolumn{1}{c}{\multirow{3}{*}{\begin{tabular}[c]{@{}c@{}}Model\\ Architecture\end{tabular}}} & \multicolumn{4}{c}{INbreast} & \multicolumn{4}{c}{CBIS-DDSM} & \multirow{3}{*}{\begin{tabular}[c]{@{}c@{}}Trainable\\ Parameters\end{tabular}} \\ 
\multicolumn{1}{c}{} & \multicolumn{2}{c}{SSDL} & \multicolumn{2}{c}{Supervised} & \multicolumn{2}{c}{SSDL} & \multicolumn{2}{c}{Supervised} &  \\ 
\multicolumn{1}{c}{} & \multicolumn{1}{c}{$\bar{x}$} & \multicolumn{1}{c}{$s$} & \multicolumn{1}{c}{$\bar{x}$} & \multicolumn{1}{c}{$s$} & \multicolumn{1}{c}{$\bar{x}$} & \multicolumn{1}{c}{$s$} & \multicolumn{1}{c}{$\bar{x}$} & \multicolumn{1}{c}{$s$} &  \\ \hline
VGG19\_bn & \textbf{0.6764} & 0.1084 & 0.6682 & 0.0770 & \textbf{0.7313} & 0.0742 & 0.5163 & 0.2826 & \multicolumn{1}{r}{139.5 Million} \\
ResNet-152 & \textbf{0.6774} & 0.1167 & 0.6767 & 0.1021 & \textbf{0.6575} & 0.1075 & 0.5857 & 0.0598 & \multicolumn{1}{r}{58.1 Million} \\
EfficientNet-b0 & \textbf{0.6512} & 0.1081 & 0.6393 & 0.0603 & \textbf{0.5982} & 0.0753 & 0.5824 & 0.0489 & \multicolumn{1}{r}{4 Million}
\end{tabular}

\end{table*}

\section{Conclusions}
\label{conclusions}

In this work we discussed the impact of using target datasets with scarce labeled data for the implementation of deep learning models for detection of malign cases using mammogram images. As presented in \cite{balki2019sample}, the determination and study of an appropriate dataset size is an open challenge. It is clear that under real-life conditions  medical imaging implementation of deep learning systems is still challenging, namely due to problems with labeled data scarcity and class-imbalance.

To tackle these challenges on the binary classification of mammograms, a combination of transfer learning from source datasets and semi-supervised learning to leverage unlabeled target data has been proposed and tested. In the experiments carried out in this work, it was found that this combination can achieve significant improvements on the classification performance of deep learning models. This surpasses the performance of models without transfer learning or without the use of unlabeled target data. The experiments depicted in this work also reveal the importance of using transfer learning from source datasets. Still, the highest yielded performance of the \gls{SSDL} model with fine-tuning have a large room for improvement. Enforcing further supervision with small labeled datasets (pixel-wise labelling of the regions of interest), with other forms of weak or self-supervision \cite{tardy2021looking} and/or domain adaptation \cite{shen2020unsupervised}, along with more complex data augmentation approaches as in \cite{domingues2018bi}, might improve the overall model performance. This must be done without raising too much the need of expensive labelling. 

The target dataset used in this work for the evaluation of the models in the classification of mammograms is made available for other interested researchers. The dataset built for this work shows real-life conditions for the deployment of a deep learning based \gls{CAD} system. Highly imbalanced data, along with the significant distribution mismatch with the source datasets are  important and frequent aspects of real-world test data for medical imaging based \gls{CAD}.

 The dissimilarity between source and target datasets was found to be significant with the use of the \gls{DeDiMs} measures. This was shown to be the case even though images from datasets can be considered as semantically and visually similar. Related to this, the choice of the source dataset was found to be an important factor in the yielded improvements in the performance of models, as well as model complexity. The measured \gls{DeDiMs} can be considered a generic and simple data quality metric, similar to the data heterogeneity metric proposed in \cite{mendez2019using}. Specific data quality metrics for deep learning models to solve medical imaging challenges is still a very under-developed topic in the literature. We plan to contribute in such  data-oriented metric development in the medical imaging analysis field in the future.  Additionally further evaluation of  model-oriented properties of  deep learning  models such as robustness and predictive uncertainty, as recommended in \cite{oala2020ml4h}, is also a future work-line to develop.

\begin{acknowledgements}
This research was partially supported by a machine allocation on Kabré supercomputer at the Costa Rica National High Technology Center. This work is partially supported by by the Ministry of Science, Innovation and Universities of Spain under grant number RTI2018-094645-B-I00, project name Automated detection with low cost hardware of unusual activities in video sequences. It is also partially supported by the Autonomous Government of Andalusia (Spain) under project UMA18-FEDERJA-084, project name Anomalous behaviour agent detection by deep learning in low cost video surveillance intelligent systems. All of them include funds from the European Regional Development Fund (ERDF). The authors thankfully acknowledge the computer resources, technical expertise and assistance provided by the SCBI (Supercomputing and Bioinformatics) center of the University of Malaga. The authors also acknowledge the funding from the Instituto de Investigación Biomédica de Málaga – IBIMA and the Universidad de M\'alaga. Lastly, the authors acknowledge the contribution and support from Luis Chavarr\'ia as a member of the executive board of the private clinic Im\'agenes M\'edicas Dr. Chavarr\'ia Estrada, for granting permission to the access and usage of mammogram images from the clinic. 
\end{acknowledgements}

\section*{Compliance with ethical standards}

\paragraph{Conflicts of interest} The authors have no conflicts of interest to declare that are relevant to the content of this article.

\paragraph{Research involving human participants} The authors declare that both this study and the data gathering process of these images comply with the Declaration of Helsinki for medical research, as this study was entirely observational, did not directly involve any human subjects. Furthermore, no identifiable human material nor data were used. The data was already acquired during regular clinical practice. Additionally, data from the INbreast and CBIS-DDSM datasets was gathered from previous third-party studies and made available for academic research.

The authors count with explicit permission from the Chavarr\'ia Clinic executive board for the usage of their images for academic purposes. Additionally, since the data was collected from patients of the clinic in 2020, it was already gathered by the beginning of this study and was ultimately provided to the research team. 

\section*{Author Biographies}

\paragraph{\textbf{Sa\'ul Calder\'on-Ram\'irez}} received his B.Sc. in computer science and his M.Sc. in electrical engineering from the University of Costa Rica, Costa Rica. He is currently a  Ph.D student at De Montfort University, U.K.

\paragraph{\textbf{Diego Murillo-Hern\'andez}} received the B.Sc. degree in computer engineering from the Costa Rica Institute of Technology, Costa Rica, in 2021. He is currently working in the private industry as a data scientist. 

\paragraph{\textbf{Kevin Rojas-Salazar}} currently an undergraduate student of Computer Engineering at the Costa Rica Institute of Technology, Costa Rica. He works  as a research assistant of deep learning applications.

\paragraph{\textbf{David Elizondo}} (Senior Member, IEEE) received his P h.D. in computer science, U.of Strasbourg, France. He is currently a professor at  De Montfort University, U.K.

\paragraph{\textbf{Shengxiang Yang}} (Senior Member, IEEE) received the Ph.D. degree from Northeastern University,  China. He is currently Director of the Centre for Computational Intelligence, De Montfort University,  U.K.

\paragraph{\textbf{Armaghan Moemeni}} received the Ph.D. degree in computer science from De Montfort University. She is currently an Assistant Professor in computer science with the University of Nottingham.

\paragraph{\textbf{Miguel Molina-Cabello}} received the Ph.D. degree in computer engineering from the University of Málaga. He works at the University of Málaga, where he holds a teaching and researching position.

\bibliographystyle{spmpsci}      
\bibliography{References}   

\end{document}